\documentclass[twocolumn,aps]{revtex4}
\usepackage{graphicx,amsfonts}
\begin{document}

\bibliographystyle{apsrev}

\title{Gauge invariance and electron
spectral functions in underdoped cuprates}
\date{\today}
\author{Walter Rantner}
\author{Xiao-Gang Wen}
\affiliation{Dept.\ of Physics, Massachusetts Institute of Technology,
Cambridge, Massachusetts 02139}
\begin{abstract}
The single particle spectral function for the normal state of underdoped high
$T_c$ cuprates is studied within the slave particle framework. We find that
the presence of a massless dynamical gauge field - a direct consequence of the
quantum order 
- explains the broad, but not totally
incoherent, line-shapes observed in experiments. 
The issue of the negative anomalous dimension of a
recently proposed gauge invariant single particle amplitude is also
considered. We show how the anomalous behavior of the single particle
amplitude can be incorporated within the slave particle approach and, thus
reinterpreted, lead to physical phenomenology.   
\newline
\newline
\end{abstract}
\maketitle

\newcommand{\dm}{\mu + 4\tilde{t}\chi}
\newcommand{\dmn}{|\mu| - 4\tilde{t}\chi}
\newcommand{\om}{i\omega_n + |\mu| - 4\tilde{t}\chi}
\newcommand{\omcon}{|\omega|-|\mu|+4\tilde{t}\chi}
\newcommand{\tbar}{\frac{tv^2+(1-t)v_b^2}{v^2v_b^2}}

\section{Introduction}

In recent years gauge theories have become a prominent tool in the description
of strongly correlated electron systems
\cite{BZA,Fourauthor,IoffeLarkin_chapter3}.  The gauge field appears because
we attempt to use fermion operators to describe excitations of
a strongly correlated state.

We know that the fermion operators are useful at low energies only when the
ground state wave function can be described by a Slater determinant. If the
ground state is a strongly correlated state, we cannot use the fermion/boson
operators to write down a free low energy effective theory.  However if we
insist on writing everything in terms of fermion/boson operators, one finds
that we can use fermion/boson operators to describe the low energy physics of
a strongly correlated state provided that gauge fields are introduced.  This
is why low energy effective theories of strongly correlated states, such as
the quantum Hall states and spin liquid states, are all gauge theories.  

Realizing that the electron operator is not a good starting point to describe
a strongly correlated state, we can rewrite the electron operator as a product
of several other operators in order to study those states.  These operators
are called parton operators (the spinon and holon operators are examples of
parton operators).  We then construct mean-field states in the Hilbert space
of partons.  After identifying the gauge structure -- the transformations
between partons that leave the electron operator unchanged, we can project the
mean-field state onto the physical Hilbert space ({\it i.e.} the gauge
invariant space) and obtain a strongly correlated electron state.  Although
the parton mean-field state is a Slater determinant, the projected state is
{\it not} and can be used to describe the correlations in certain strongly
correlated states.  This procedure, in its general form, is called projective
construction, which is a generalization of the slave-boson
approach\cite{BZA,WenLee,su2_chapter3,Wsrvb_chapter3}.  The
generic projective construction and the associated gauge structure was
discussed in detail for quantum Hall states in \cite{Wpc_chapter3}. The resulting
effective theory arising from the projective construction naturally contains a
gauge structure. In this paper, we will use the projective-construction/
slave-boson-approach to study  the electron spectral function in underdoped
high $T_c$ superconductors.

 One convenient way to describe dynamical gauge theories is via the
path integral approach.  Traditionally, however, the formalism which allows
for most direct comparison with experimental results is based on Green's
functions and diagrammatic techniques. This turns out to be a kind of nuisance
since the standard expression for the single particle Green's function (the
cornerstone of diagrammatic perturbation theory) is not gauge invariant. The
lack of gauge invariance makes it hard to extract any meaningful quantitative
results based on these Green's functions. 
 
In the first part of the paper we present a calculation of a
particular gauge invariant Green's function for  2+1 dimensional massless
Dirac particles coupled to a dynamical gauge field. The motivation for this
analysis arises from an effective continuum description - the Algebraic Spin
Liquid (ASL)\cite{ASL, CHI}, which we believe underlies the strange
phenomenology of the normal state of underdoped high temperature
superconductors. Within this continuum model, which is rooted in the
slave-boson approach to strongly correlated systems, the spin degrees
of freedom are carried by massless Dirac particles called spinons.  Our
early attempt to interpret the resulting spinon propagator as describing the
propagation of the physical particle is however thwarted by the appearance of
a negative anomalous dimension for the particular gauge invariant spinon
amplitude at the ASL fixed point. On reinterpreting the proposed amplitude as
representing a two-particle spinon-holon propagator, we find that the main
effect of the gauge fluctuation is to bind the two particles (via its effect
on the vertex) to a more coherent entity. Thus we are led to the  heuristic
description of the physical electron propagator presented in the last section of
the paper where in addition to the spinon contribution we include the bosonic
holon contribution.
Not surprisingly the
resulting spectrum has no quasiparticle peak, however it is also not totally
incoherent once the gauge fluctuations are included.  The emerging picture for
the electron propagation is thus one of spinons and holons (separate at
the mean field level) whose attractive interaction mediated via the gauge
field leads to the formation of a more coherent structure on top of the
incoherent mean-field background \cite{Laughlin97}.  In analyzing this problem
we have been guided by a first quantized path integral approach to the
physical Green's function which makes the gauge dependence of the amplitude
particularly transparent. 

\section{Path Integral formulation}\label{section2}

 Before plunging into the second quantized formalism let us look at
the problem in the light of single particle quantum mechanics.
Our starting point is the following definition for a gauge-invariant Green's
function in first quantized notation
\begin{equation}
 G(\vec{x}) =
\int Da Dx e^{-\int_0^{\vec{x}} d\tau (L[x(\tau),a]+ L[a])}
e^{-i\int_\mid a_{\mu} dx_{\mu}}
\end{equation}
where $\mid$ is the straight line
connecting $0$ to $\vec{x}$,
$\mu = 0..2$ and the metric is Euclidean. The action for the gauge
field is given by 
\begin{equation}\label{gaugeaction}
S[a_{\mu}] = \frac{1}{2}\int \frac{d^{3}q}{(2\pi)^{3}}
a_{\mu}(\vec{q})\frac{1}{4}\sqrt{q^2}(\delta_{\mu \nu} -
\frac{q_{\mu}q_{\nu}}{q^{2}})a_{\nu}(-\vec{q})
\end{equation}
which is obtained by integrating out massless Dirac fermions (the
spinons) [See Eq. (\ref{polarisation})].
The action for the particle moving in an external gauge field can be taken to
have the following form
\begin{equation}
\int_0^{\vec{x}} d\tau L[x(\tau), a]=
\int_0^{\vec{x}} d\tau L[x(\tau)] + \int_{\wr} a_{\mu} dx_{\mu}
\end{equation}
where $\int_{\wr}$ indicates a path from $0$ to $\vec{x}$ described by 
$x(\tau)$.  $L[x(\tau)]$ remains unspecified since we will
be concerned with different types of particles. 
Putting the above together we
obtain the following expression for the Green's function
\begin{eqnarray}
 G(\vec{x}) =
\int Da Dx e^{-\int_0^{\vec{x}} d\tau (L[x(\tau)]+ L[a])}
e^{i\int_{\mid}a_{\mu} dx_{\mu}-i\int_{\wr}a_{\mu} dx_{\mu}}
\end{eqnarray}
This expression exhibits the gauge dependence in a particularly appealing form
where the line integrals close up into a contour, of which one side is the
propagating path of the particle and other side the straight-line path 
needed to make the gauge
invariance manifest. Via Stokes theorem we see that the sum of the line
integrals is nothing but the flux going through the contour.   So far
our expressions have full rotational symmetry in order to perform the
calculation we now chose to parameterize the particle world-lines by their
$x_{0}=\tau$ coordinate. Hence $\int a_{\mu}dx_{\mu}$ goes over to $\int ({\bf
a(x)}\frac{d{\bf x}}{d\tau} + a_{0})d\tau$ where ${\bf a},{\bf x}$ denote the
2 spatial components of the corresponding 3-vectors.  In order to simplify our
analysis further we consider the propagation of the particle only along the
time direction i.e. $G(\mathbf{0},\tau)$.  Considering the propagation between
equal spatial coordinates we have no contribution from the 2-velocity
component along the straight line path which simplifies this part of the phase
factor to $\int_{\mid} a_{0}d\tau$.   Next we want to perform the
average over the gauge fluctuations which is straightforward given the
Gaussian weight (\ref{gaugeaction}). Reading off the gauge propagator
\begin{displaymath}
D\left(a_{\mu}(\vec{q})a_{\nu}(\vec{k})\right) =
\frac{1}{(2\pi)^3}\delta(\vec{q}+\vec{k})\left[\frac{1}{4}\sqrt{\vec{q}^2}(\delta_{\mu \nu} - \frac{q_{\mu}q_{\nu}}{q^{2}})\right]^{-1}
\end{displaymath}
In order to invert the polarization we chose the $a_{0}=0$ gauge which yields
\begin{displaymath}
D\left(a_{\mu}(\vec{q})a_{\nu}(\vec{k})\right) =
(2\pi)^3\delta(\vec{q}+\vec{k})\delta_{\mu,i}\delta_{\nu,j}\left(\delta_{i,j} +
\frac{q_{i}q_{j}}{q_{0}^2}\right)\frac{4}{\sqrt{\vec{q}^2}}
\end{displaymath}
where $i,j=1,2$ denote the spatial components of $\vec{q}$.   Note
particularly that in choosing the $a_{0}=0$ gauge we got rid of the remaining
contribution of the straight line path to the phase factor in the expression
for the Green's function.   Now we perform the average over the gauge
fluctuations
\begin{eqnarray}
\langle exp(-i\int_{0}^{\tau}{\bf a}\dot{{\bf x}} d\tau) \rangle_{{\bf a}} =
\quad \quad \quad \quad  \quad \quad \quad \quad \quad \quad  \nonumber\\
exp\Big( - \frac{1}{2}\int_{0}^{\tau} \int_{0}^{\tau} \dot{{\bf
x}}(\tau)\dot{{\bf x}}'(\tau') d\tau d\tau'
\sum_{\vec{q},\vec{k}}e^{i\vec{q}\vec{x}}e^{i\vec{k}\vec{x}'}\langle {\bf
a}(\vec{q}){\bf a}(\vec{k}) \rangle_{{\bf a}}\Big)\nonumber
\end{eqnarray}
The calculation is straightforward albeit some care has to be taken with the
boundary conditions. The averaging results in
\begin{eqnarray}\label{effectiveinteraction}
exp\left( - \frac{1}{4\pi^2}\int_{0}^{\tau} \int_{0}^{\tau} \theta(\tau-\tau')d\tau
d\tau'\times \right.\nonumber \\
\left[ \frac{\dot{\bf x}(\tau)\dot{\bf x}'(\tau') + 1}{[{\bf x}-{\bf x}']^{2} +
(\tau-\tau')^{2}} - \frac{1}{{\bf x}^{2}+(\tau-\tau')^2}\right. \nonumber \\
\left. \left.- \frac{1}{{\bf x}'^{2}+(\tau-\tau')^2} + \frac{1}{(\tau-\tau')^{2}}\right]\right)
\end{eqnarray}
The last three terms in the above expression (\ref{effectiveinteraction})
arise from the boundary conditions and thus explicitly depend on the direction
of propagation. In order to extract the dependence on
$\vec{x}_{f}-\vec{x}_{i}$ explicitly we can now exploit the rotational
invariance in the Euclidean formulation to re-express the above in vector
notation.
\begin{eqnarray}\label{rotationalinvariantinteraction}
exp\left( -
\frac{1}{4\pi^2}\int_{\vec{x}_{i}\cdot\vec{n}}^{\vec{x}_{f}\cdot\vec{n}}
\int_{\vec{x}_{i}\cdot\vec{n}}^{\vec{x}_{f}\cdot\vec{n}}
\theta(\vec{x}\cdot\vec{n} - \vec{x}'\cdot\vec{n}) \right.  \nonumber \\
\left. \Big[ \frac{d\vec{x}\cdot d\vec{x}'}{(\vec{x} - \vec{x}')^{2}} -
\frac{d\vec{x}\cdot\vec{n}
d\vec{x}'\cdot\vec{n}}{[\vec{x}-(\vec{x}'\cdot\vec{n})\vec{n}]^{2}} \right.
\nonumber \\
\left. - \frac{d\vec{x}\cdot\vec{n}
d\vec{x}'\cdot\vec{n}}{[\vec{x}'-(\vec{x}\cdot\vec{n})\vec{n}]^{2}} +
\frac{d\vec{x}\cdot\vec{n}
d\vec{x}'\cdot\vec{n}}{[(\vec{x}\cdot\vec{n})\vec{n}
-(\vec{x}'\cdot\vec{n})\vec{n}]^{2}}\Big]\right)
\end{eqnarray}
with $\vec{n} = \frac{\vec{x}_{f}-\vec{x}_{i}}{|\vec{x}_{f}-\vec{x}_{i}|}$ the
unit vector along the classical straight line path.   Expressions
(\ref{effectiveinteraction}) and (\ref{rotationalinvariantinteraction})
exhibit nicely the retarded nature of the effective interaction. In principle,
to obtain the propagator $G(\vec x)$,
we would now have to perform the average over the trajectories with the free
particle action appropriate for the particle in question.  For a
\emph{massive} particle we can neglect the spatial part of the retardation
since it is down by a factor of the speed of ``light'' (the spinon velocity)
which we have set equal to
one. Hence we can see that only the term proportional to $\dot{{\bf
x}}(t)\dot{{\bf x}}'(t')$ gives a non-vanishing contribution whose effect is
to renormalize the mass of the particle.   For a massless
relativistic particle however both spatial and temporal retardation are
equally important and the argument above cannot be applied.  We find that the
integrals in (\ref{effectiveinteraction}) are all dimensionless. In the large
$\tau$ limit, those integrals have the form $c_1 \frac{\tau}{l_0} + c_2 \ln
\frac{\tau}{l_0}$, where $l_0$ is the short distance cut-off scale.  The $c_1$
term corresponds to mass generation. Within second quantization, as we will
discuss next, the mass term cannot be generated if we regularize our theory at
short distances in a way consistent with the underlying lattice symmetries +
gauge structure \cite{Wqo}. Thus we can set the regularization dependent term
$c_1$ to zero. The true leading term has a form $c_2 \ln \tau$ ($c_2$ is
regularization independent).  Those contributions can only change the exponent
of the free Green's function: $1/\tau^2 \to  1/\tau^{(2-\alpha)}$.  The effect
of the gauge interaction is to modify the exponent in the algebraic decay of
the Green's function, which means that the gauge interaction is a marginal
perturbation.

\section{Massless Dirac spinors}\label{fgreenfunction}

 Having used the first quantized version of the path integral to
guide our intuition we will henceforth return to field theory and in
particular to the problem of massless Dirac spinors coupled to a U(1) gauge
field.  The Euclidean model which we are concerned with is the following
\begin{eqnarray}\label{QED3}
Z_{\Psi} = \int D\bar{\Psi}D\Psi 
  exp\Big( -\int d^{3}x \sum_{\sigma=1}^N
\bar{\Psi}_{\sigma}(\partial_{\mu}-ia_{\mu})\gamma_{\mu}\Psi_{\sigma}\Big) \nonumber \\
\end{eqnarray}
with the dynamics for the gauge field given by
\begin{eqnarray}\label{polarisation}
Z_{a}&=&\int Da_{\mu}\exp\Big( -\frac{1}{2}\int\frac{d^3q}{(2\pi)^3}a_{\mu}
(\vec{q})\Pi_{\mu\nu}a_{\nu}(-\vec{q})\Big) \nonumber \\
\Pi_{\mu\nu}&=&\frac{N}{8}\sqrt{\vec{q}^2}\Big(\delta_{\mu\nu}
- \frac{q_{\mu}q_{\nu}}{\vec{q}^{2}}\Big)
\end{eqnarray}
 The Fermi field $\Psi_{\sigma}$ is a $4\times 1$ spinor
$\psi^{\dag}_{\sigma} \equiv (\psi^{\dag}_{\sigma,1},\psi^{\dag}_{\sigma,2})$
with $\psi_{\sigma,1}=$ $ f_{\sigma 1 e} \choose f_{\sigma 1 o}$
,$\psi_{\sigma,2} =$ $ f_{\sigma 2 o} \choose f_{\sigma 2 e}$. The notation
(1,2,e,o) arises naturally when the theory (\ref{QED3}) is derived as the low
energy effective description of the so called $\pi$-flux phase of Hubbard type
lattice models \cite{MarstonAffleck}. There 1,2 denotes two types of Fermi
points and e,o denote even and odd lattice sites respectively.  Notice that we
have generalized to N fermion species in order to have a more controlled
perturbation theory (N=2 in the physical case).  The $4\times4$ $\gamma_{\mu}$
matrices form a representation of the Dirac algebra
$\{\gamma_{\mu},\gamma_{\nu}\}=2\delta_{\mu\nu}$ ($\mu,\nu = 0,1,2$) and are
taken to be
\begin{eqnarray}
\gamma_{0}&=&\pmatrix{\sigma_{3}&0\cr0&-\sigma_{3}\cr}, \quad
\gamma_{1}=\pmatrix{\sigma_{2}&0\cr0&-\sigma_{2}\cr}, \\
\gamma_{2}&=&\pmatrix{\sigma_{1}&0\cr0&-\sigma_{1}\cr} \nonumber
\end{eqnarray}
with $\sigma_{\mu}$ the Pauli matrices.  Finally note that
$\bar{\Psi}_{\sigma} \equiv \Psi^{\dag}_{\sigma} \gamma_{0}$.

\begin{figure}[tb]
\begin{center}
\includegraphics[width=70mm]{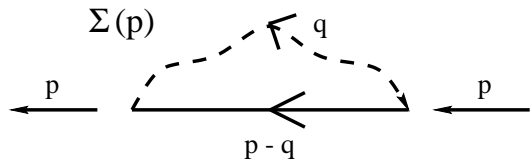}
\caption{Exchange diagram used in the calculation of the self-energy}
\label{Exchangediagram}
\end{center}
\end{figure}

>From the path integral discussion above we have learned a way to calculate a
particular gauge invariant Green's function for the Fermi fields.  Note that a
certain choice for the gauge has to be made before we can invert the
polarization operator to obtain the gauge propagator.  In order to get rid of
the contribution from the straight line path in $G$, we can fix the gauge in
such a way as to make it vanish along that very line. In this particular
gauge, $G$ has a form of a normal one-particle Green's function and can be
calculated using the usual method through the self energy:
\begin{displaymath}
G(\vec{p}) = G_{0} (\vec{p})+ G_{0}(\vec{p})\left(-\Sigma(\vec{p})\right)G_{0} (\vec{p})+ \cdots
\end{displaymath}
where $\Sigma$ to lowest order in $1/N$ is the exchange diagram depicted in
Fig.[\ref{Exchangediagram}].  However, the above trick only works for one
particular direction $\vec{x}$ in real space.  For different directions, we
need to choose different gauges.   We can put this into a more
precise mathematical form.  By choosing a gauge $\hat{n}\cdot \vec{a} =0$, we
can calculate the self energy and the corresponding Green's function (which is
denoted as $G_{\hat{n}}(\vec{x})$) through diagrams.  However,
$G_{\hat{n}}(\vec{x})$ is not equal to the gauge invariant Green's function
$G(\vec{x})$.  But the two Green's functions are closely related:
\begin{equation}
 G(\vec{x}) = G_{\vec{n}}(\vec{x})|_{\hat{n}=\vec{x}/|\vec{x}|}
\end{equation}
In the following, we will calculate $G$ through calculating $G_{\hat{n}=\hat
\tau}$.

To leading order in the gauge field fluctuations the fermion self-energy is
\begin{displaymath}
-\Sigma(\vec{p}) = \int \frac{d^3 q}{(2\pi)^{3}}
i\gamma^{\mu}G_{0}(\vec{p}-\vec{q})i\gamma^{\nu}D_{\mu\nu}(\vec{q})
\end{displaymath}
where $G_{0} = \frac{-ik_{\nu}\gamma^{\nu}}{k^{2}}$ is the free fermion
Green's function.  The sign convention for the self energy has been chosen
such that
\begin{displaymath}
G^{-1}(k) = ik_{\mu}\gamma^{\mu} + \Sigma(k)
\end{displaymath}
To invert the gauge polarization, we will choose the $a_{0} = 0$ gauge.  This
implies that our calculation gives a correct result for the equal-space
propagation only. Employing relativistic invariance however, we can
immediately extend the result to all of real space.   With the above
gauge choice 
\begin{displaymath}
D_{\mu\nu}(\vec{q}) =
\frac{8}{N}\frac{1}{\sqrt{\vec{q}^{2}}}\delta_{\mu i}\delta_{\nu j} \Big(
\delta_{i j} + \frac{q_{i}q_{j}}{q_{0}^{2}}\Big)
\end{displaymath}
and we arrive at the following expression for the self-energy at T=0

\begin{eqnarray}\label{self-energy}
\frac{N}{8}&i&\Sigma(\vec{p}) = \\
& &\gamma^{0}\int \frac{d^{3}q}{(2\pi)^{3}}\left[
\frac{(q_{0}-p_{0}){\bf q}^{2} + 2 (q_{0}-p_{0})
q_{0}^{2}}{\sqrt{\vec{q}^{2}}(\vec{p}-\vec{q})^{2}q_{0}^{2}}\right] 
\nonumber \\
&+&\gamma^{x}\int \frac{d^{3}q}{(2\pi)^{3}}\left[
\frac{(p_{x}-q_{x})(q_{x}^{2}-q_{y}^{2}) +
2q_{x}q_{y}(p_{y}-q_{y})}{\sqrt{\vec{q}^{2}}
(\vec{p}-\vec{q})^{2}q_{0}^{2}}\right]\nonumber \\
&+&\gamma^{y}\int \frac{d^{3}q}{(2\pi)^{3}}\left[
\frac{(p_{y}-q_{y})(q_{y}^{2}-q_{x}^{2}) +
2q_{x}q_{y}(p_{x}-q_{x})}{\sqrt{\vec{q}^{2}}
(\vec{p}-\vec{q})^{2}q_{0}^{2}}\right]\nonumber
\end{eqnarray}

In order to evaluate the self energy we need to take care of both the UV
divergences usually encountered in loop integrations and the singularities
arising from the $1/q_0^2$ terms which are a hallmark of the incomplete gauge
fixing in the temporal axial gauge.  Details of this calculation are presented
in the appendix.   We find for the self-energy
\begin{eqnarray}\label{sigma_a0}
\Sigma_{0}(\vec{p}) &=& -i\frac{16}{N}\frac{1}{3\pi^2}p_0\gamma^0 ln\left(\frac{\Lambda}{|\vec{p}|}\right)^2 \\ \nonumber
\Sigma_{i}(\vec{p}) &=& -i\frac{8}{N}\frac{1}{3\pi^2}p_i\gamma^i ln\left(\frac{\Lambda}{|\vec{p}|}\right)^2
\end{eqnarray}

Had the self-energy taken the form
\begin{eqnarray}\label{sigma_a0S}
\Sigma_{0}(\vec{p}) &=& -i\gamma_{\Psi} p_0\gamma^0 ln\left(\frac{\Lambda}{|\vec{p}|}\right)^2 \\ \nonumber
\Sigma_{i}(\vec{p}) &=& -i\gamma_{\Psi} p_i\gamma^i ln\left(\frac{\Lambda}{|\vec{p}|}\right)^2
\end{eqnarray}
we could read off the anomalous dimension of the
fermion field and found the Green's function $G_{\hat \tau}(x)$
\begin{equation}
G_{\hat \tau}(\vec{x}) \propto
\frac{x_{\mu}\gamma^{\mu}}{\vec{x}^{3+2\gamma_{\Psi}}}
\end{equation}
Since the above $G_{\hat \tau}$ has no explicit $\hat\tau$ dependence, we find
\begin{equation}
G(\vec{x}) = G_{\hat \tau}(\vec{x}) \propto
\frac{x_{\mu}\gamma^{\mu}}{\vec{x}^{3+2\gamma_{\Psi}}}
\end{equation}

However, the coefficients of the two terms 
in Eq. (\ref{sigma_a0}) are not equal and the above
calculation of $G(\vec{x})$ is not valid. But noting that the coefficients
in Eq. (\ref{sigma_a0}) have the same sign - corresponding to a negative
$\gamma_{\Psi}$ - this allows us to conclude that $G_{\hat \tau}(\tau)$ has the form
\begin{equation}
G_{\hat \tau}(\tau) \propto
\frac{\tau\gamma^0}{\tau^{3+2\gamma_{\Psi}}},\ \ \ \
-\frac{16}{N}\frac{1}{3\pi^2} \leq \gamma_{\Psi} \leq -\frac{8}{N}\frac{1}{3\pi^2}
\end{equation}
We see that the effect of gauge interactions is to make the propagator decay
slower. In the path integral picture, the gauge interactions bring the
particle path closer to the straight-line path so the area enclosed by the
two paths is smaller. Since the particle path is closer to the straight-line path in the $\tau$ direction, this leads us to believe that the second $p_i$
term in Eq. (\ref{sigma_a0}) is not important compared to the first $p_0$
term. Hence
\begin{equation}
 \gamma_{\Psi} = -\frac{16}{N}\frac{1}{3\pi^2}
\end{equation}
This gives us
\begin{equation}
\label{Galpha}
G \propto \frac{x_\mu\gamma^\mu}{x^{3+2\gamma_{\Psi}}},\ \ \ \
\gamma_{\Psi} = -\frac{16}{N}\frac{1}{3\pi^2}
\end{equation}

We would like to mention that
the renormalization program within the temporal axial
gauge is not  established (see Ref. \cite{Bassetto} for a discussion on
renormalization within non-covariant gauges).   
To check and to confirm the result Eq. (\ref{Galpha}),
we follow a different route where we consider in addition to the
massless fermions coupled to the gauge field a heavy boson field whose
``current'' will take care of the straight line path in the equal space
correlater. This type of formulation is borrowed from heavy quark effective
theory approaches to QCD \cite{Grozin}.  The task at hand is thus a
calculation of the wavefunction renormalization of the boson fermion bilinear
$j = \bar{\Psi}b$ with the following Lagrangian
\begin{displaymath}
\mathcal{L} = \bar{\Psi}_{\sigma}(\partial_{\mu}\gamma^{\mu} + ia_{\mu}\gamma^{\mu})\Psi_{\sigma} + b^{*}(\partial_{0} + ia_{0})b + \mathcal{L}_{a}
\end{displaymath}
In order to calculate the wavefunction renormalization of the above composite
we need the field strength renormalizations $Z_{\Psi}, Z_{b}$ for both the
fermion and the boson field as well as the vertex renormalizations
$Z_{\Gamma}$.
\begin{figure}[tb]
\begin{center}
\includegraphics[width=70mm]{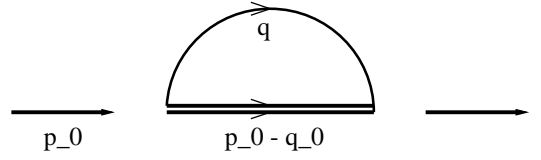}
\caption{Exchange diagram used in the calculation of the boson self-energy}
\label{Selfenergy_boson}
\end{center}
\end{figure}
To obtain the boson wavefunction renormalization we calculate its self energy
due to the gauge field fluctuations Fig[\ref{Selfenergy_boson}]. As can be
seen from the above Lagrangian, the boson being static only couples to the
Coulomb field. Thus we arrive at the following expression for the boson self
energy.
\begin{equation}\label{boson_sigma}
-\Sigma_{b}(p_0) = \int\frac{d^3q}{(2\pi)^2}\frac{i\gamma_{0}D_{00}(\vec{q})i\gamma_{0}}{i(p_{0}-q_{0})}
\end{equation}
whose divergent part evaluates (see appendix) to
\begin{equation}
i\Sigma_{b}(p_0) = p_0 \frac{8}{\pi^2N}\frac{1}{3-d}
\end{equation}
where the sign convention for the self energy has been chosen in accord with
the one for the fermion field mentioned above.

 Thus we find for the wavefunction renormalization the minimal
\begin{displaymath}
G_b(p_0) = \frac{1}{ip_0 + \Sigma(p_{0})} = \frac{Z_b}{ip_0}
\quad Z_b = 1+\frac{8}{\pi^2N}\frac{1}{3-d}
\end{displaymath} 
This result has been obtained in the Landau gauge.  The fermion wavefunction
renormalization can be obtained similarly from the self energy  and reads in
the Landau gauge
\begin{equation}
Z_{\Psi} = 1 + \frac{8}{3\pi^2N}\frac{1}{3-d}
\end{equation} 

\begin{figure}[tb]
\begin{center}
\includegraphics[width=30mm]{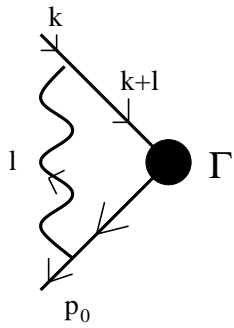}
\caption{Boson-Fermion vertex with the $p_0$ flowing along the static boson line}
\label{vertex_chapter3}
\end{center}
\end{figure}
Finally we need to obtain the vertex renormalization. From
Fig.[\ref{vertex_chapter3}] we can read off the following expression for the
vertex correction
\begin{displaymath}
\int\frac{d^3l}{(2\pi)^3}\frac{i\gamma^{\mu}(-i)\gamma^{\epsilon}(k+l)_{\epsilon}\openone \delta_{0\nu}i D_{\mu\nu}(l)}{(k+l)^2 i(p_0+l_0)}
\end{displaymath}
where $\openone$ is the fermion-boson vertex whose renormalization is sought.
By simple powercounting we see that the loop integration is only
logarithmically divergent and hence we can evaluate it by setting all external
moment equal to zero which leaves us with
\begin{eqnarray*}
\openone\int\frac{d^3 l}{(2\pi)^3}\left[ \frac{\gamma_0 \gamma^{\epsilon}l_{\epsilon}}{l^3l_0} - \frac{\gamma^{\mu}\gamma^{\nu}l_{\mu}l_x{\nu}}{l^5}\right] \\
= \openone\int\frac{d^3 l}{(2\pi)^3}\left[\frac{1}{l^3} - \frac{l^2}{l^5}\right] = 0
\end{eqnarray*}
where the first term was simplified by the fact that only the $\gamma^0 l_0$
term survives integration over spatial momenta.  Thus we conclude that within
the Landau gauge the vertex doesn't renormalize to one loop order.

 We now have all the ingredients to calculate the wave function
renormalization of the fermion boson bilinear in the minimal form
\begin{equation}
Z_{j} = Z_b^{1/2}Z_{\Psi}^{1/2}Z_{\Gamma} = 1 + \frac{16}{N}\frac{1}{3\pi^2}\frac{1}{3-d}
\end{equation}
This result is gauge invariant. From the above discussion we can see that
within the axial temporal gauge there is no Coulombic interaction and hence
both the vertex correction and the boson wavefunction renormalization are
absent which would suggest that 
\begin{displaymath}
Z_{\Psi}(a_{0} = 0) = Z_{j}^2 = 1 + \frac{16}{N}\frac{1}{3\pi^2}\frac{2}{3-d}
\end{displaymath} which yields 
\begin{equation}
\gamma_{\Psi} = -\frac{16}{N}\frac{1}{3\pi^2}
\end{equation}
for the anomalous dimension of the fermion field.

 An important point to note here is the fact that the gauge
fluctuations do not generate a mass or a chemical-potential term
perturbatively.  In a diagram calculation, a regularization dependent
mass/chemical-potential term actually appears in the self-energy. We have set
such a regularization dependent mass/chemical-potential term equal to zero.
In the next section we will see that our Dirac-$U(1)$ gauge system is the low
energy effective description of the so called staggered flux (sF) phase (the
mean-field phase corresponding to the pseudogap regime within the $SU(2)$
slave boson theory).  The lattice sF phase provides a short distance
regularization.  Now the question is whether such a regularization generates a
regularization dependent mass/chemical-potential term or not.  One way to
answer the above question is through a symmetry argument.  In Ref.
\cite{Wqo}, it is shown that any change of the mean-field sF ansatz cannot
generate a mass term or chemical potential term if the change does not break
any symmetries and does not break the $U(1)$ gauge structure.
This implies that if the regularization is consistent with
the underlying lattice symmetries and does not break the $U(1)$ gauge
structure, then the regularization cannot generate any mass/chemical-potential
terms. This is the reason why we can set the regularization dependent
mass/chemical-potential term equal to zero. We see that the massless Dirac
fermion is protected by the lattice symmetries and by the $U(1)$ gauge
structure (or more precisely, protected by the quantum order\cite{Wqo}).  Thus
what we really have learned from the diagram calculation is that the mass or
chemical-potential is not generated spontaneously from the infrared
divergences (at the perturbative level).  Combining the symmetry consideration
for the high energy cutoff and the diagram calculation for the infrared, we
find that no mass or chemical-potential is generated from the self energy
corrections to the fermion Green's function.
In contrast, the first quantized path integral analysis does not know about
the underlying lattice symmetries and, as a result
allows for corrections to the mass or chemical-potential due to dressing with
the short-distance gauge fluctuations (see discussion in section
\ref{section2}).  We would like to mention that the Dirac-$U(1)$ gauge system
was studied in detail in Ref. \cite{QED3_chapter3} using a different
gauge fixing. It was shown that the mass/chemical-potential term is not
generated perturbatively to all orders in a $1/N$ expansion.
Furthermore note that since the conserved current (that couples to
$a_{\mu}$) cannot have any anomalous dimension, the gauge interaction is an
{\em exact} marginal perturbation.

 Hence we can take the anomalous dimension seriously and obtain the
full Green's function in the form
\begin{equation}\label{FullGreen}
G(\vec{k}) = -iC\frac{k_{\mu}\gamma^{\mu}}{\vec{k}^{2-2\gamma_{\Psi}}} 
\quad\quad
\gamma_{\Psi} = - \frac{16}{N}\frac{1}{3\pi^{2}}
\end{equation}
where $C$ is determined by the energy range over which the effective theory is
supposed to be valid.  Comparing this dressed propagator with the free fermion
Green's function $G_{0} = \frac{-ik_{\nu}\gamma^{\nu}}{k^{2}}$ we see that the
inclusion of the gauge fluctuations has destroyed the coherent quasiparticle
pole.

 At this point a couple of comments are in order. So far we have not addressed
the fact that the sign of the anomalous dimension (leading to a slower decay
of the single particle propagator in real space) is highly counter intuitive.
Fermion-fermion interaction is \emph{normally} expected to result in a
suppression of the single particle propagator - not an enhancement. This fact
combined with wishful thinking has led to the QED3 picture of the single
particle spectrum proposed in Ref. \cite{ASL}. This paper initiated a
debate about the true value and sign of the gauge invariant anomalous
dimension in massless QED3 which is ongoing \cite{QED3picture,Khveshchenko}.
The resolution of the question of the {\bf right} gauge invariant ``dressed
charge'' propagator (as Khveshchenko puts it \cite{Khveshchenko}) is an
interesting and important question for QED3 and beyond the scope of the
current analysis. However, what we must conclude from the above stated
calculation is the fact that the gauge invariant object which was studied here
- a massless Dirac fermion coupled to an infinitely heavy particle (the
straight-line path above) at the ASL fixed point  has a negative anomalous
dimension. In hindsight this can be understood as a sign of confinement - the
effect of the gauge fluctuations is the tendency to form a bound state between
the two particles.  This picture of confinement is one of central importance
to the question of spin charge separation. 

In the following we will outline a calculation of the electron
spectral function which is more directly rooted in the SU(2) approach of Wen
and Lee \cite{WenLee}. Here we include the effect of the gauge fluctuations -
the confining tendency, gleaned from the above analysis - more heuristically.

\section{Confinement by hand}

 It was mentioned above that the problem of massless Dirac particles
coupled to a gauge field is intimately related to a possible model description
of the copper oxide planes in the  high $T_{c}$ superconductors
\cite{MarstonAffleck}. To keep the discussion reasonably self-contained we
will first relate some of the details of the mean field flux state as it
appears in the SU(2) slave boson formulation \cite{Fourauthor}.

 The slave boson theories approach the copper oxide problem from the
insulating side; restricting the single particle Hilbert space on each Copper
site to three states with the doubly occupied one prohibited via the use of a
Lagrangian multiplier field. Within a path integral implementation of the
above mentioned constraint it is the Lagrange multiplier that gets promoted to
the temporal component of a gauge field.  The starting point for the
investigation is the t-J Hamiltonian which is believed to contain the
essential physics and exhibits nicely the competition of kinetic versus local
spin fluctuations.  Within the SU(2) formulation the physical electron
operator is represented as follows:
\begin{eqnarray}\label{SU2rep}
c_{\uparrow i}=\frac{1}{\sqrt{2}}h^{\dag}_{i}\psi_{\uparrow i} =
\frac{1}{\sqrt{2}}\big(b^{\dag}_{1i}f_{\uparrow
i}+b^{\dag}_{2i}f^{\dag}_{\downarrow i}\big) \\ \nonumber
c_{\downarrow i}=\frac{1}{\sqrt{2}}h^{\dag}_{i}\psi_{\downarrow i} =
\frac{1}{\sqrt{2}}\big(b^{\dag}_{1i}f_{\downarrow
i}-b^{\dag}_{2i}f^{\dag}_{\uparrow i}\big) \\ \nonumber
\end{eqnarray}
where the following SU(2) doublets were introduced
\begin{eqnarray}
\psi_{\uparrow i} = {f_{\uparrow i} \choose f^{\dag}_{\downarrow i}}, \quad
\psi_{\downarrow i} = {f_{\downarrow i} \choose -f^{\dag}_{\uparrow i}},
\quad
h_{i} = {b_{1i} \choose b_{2i}} \nonumber
\end{eqnarray}
The $\psi_{\uparrow i}, \psi_{\downarrow i}$ are the two fermion fields
representing the destruction of a spin up and spin down on site i respectively
and $h_{i}$ is the doublet of bosonic fields keeping track of the doped holes.
The procedure is now to put this representation into the t-J Hamiltonian
\begin{displaymath}
H = P\sum_{(ij)}\big[ J(\vec{S}_{i}\cdot\vec{S}_{j} - \frac{1}{4}n_{i}n_{j})
- t(c^{\dag}_{\sigma i}c_{\sigma j} + h.c.)\big]P
\end{displaymath}
which on performing a Hubbard-Stratonovich transformation to the appropriate
bosonic bond variables yields the following partition function
\begin{eqnarray}\label{SU2_partition}
Z &=& \int DhDh^{\dag}D\psi^{\dag} D\psi
D\vec{a}_{0}DUe^{-\int_{0}^{\beta}L} \\
L &=& \frac{\tilde{J}}{2}\sum_{<ij>}Tr[U^{\dag}_{ij}U_{ij}] +
\frac{1}{2}\sum_{i,j,\sigma}\psi^{\dag}_{\sigma
i}(\partial_{\tau}\delta_{ij}+\tilde{J}U_{ij})\psi_{\sigma j} \nonumber \\
 &+& \sum_{il}a^{l}_{0i}\big(\frac{1}{2}\psi^{\dag}_{\sigma
i}\tau^{l}\psi_{\sigma i} + h^{\dag}_{i}\tau^{l}h_{i}\big) \nonumber \\
 &+&
\sum_{ij}h^{\dag}_{i}\big((\partial_{\tau}-\mu)\delta_{ij}+\tilde{t}U_{ij}\big)h_{j}
\end{eqnarray}
The $\vec{a}_{0}$ fluctuations incorporate the projection to the space of
SU(2) singlets. Furthermore note that $\tilde{J}=3J/8, \quad \tilde{t}=t/2$
\cite{UbbensLee} and the matrix $U_{ij}$ in the form
\begin{displaymath}
U_{ij} = \pmatrix{-\chi_{ij}^{*}&\Delta_{ij} \cr
\Delta^{*}_{ij}&\chi_{ij}\cr}
\end{displaymath}
contains the Hubbard-Stratonovich fields which classify the part of the phase
diagram we are looking at. The mean-field phase diagram (see Introduction) is
found by minimizing the free energy for a given number of particles with
respect to the bond variables $U_{ij}$.  The phase which will be of interest
to us is the so called staggered flux (sF)phase which can be represented as
\begin{eqnarray}\label{sflux}
U_{i,i+\hat{x}} &=& -\tau^{3}\chi - i(-)^{i_{x}+i_{y}}\Delta \nonumber \\
U_{i,i+\hat{y}} &=& -\tau^{3}\chi + i(-)^{i_{x}+i_{y}}\Delta
\end{eqnarray}
The slave boson representation results in the following expression for the
physical electron Green's function 
\begin{eqnarray*}
G(\vec{r},\tau) &=& -\langle
T_{\tau}\big(c_{\uparrow}(\vec{r},\tau)c^{\dag}_{\uparrow}(\vec{0},0)\big)\rangle \nonumber \\
  &=& -\frac{1}{2} \langle T_{\tau}\big(
(h^{\dag}(\vec{r},\tau)\psi_{\uparrow}(\vec{r},\tau)\psi^{\dag}_{\uparrow}(\vec{0},0)h(\vec{0},0)\big) \rangle
\end{eqnarray*}
In the sF state we don't have any anomalous fermion-fermion pairing (both
$\tau^3$ and $\openone$ are diagonal in isospin space) and hence this
simplifies to
\begin{eqnarray*}
G(\vec{r},\tau) = &-&\frac{1}{2}\Big( \langle
b^{\dag}_{1}(\vec{r},\tau)b_{1}(\vec{0},0)f_{\uparrow}(\vec{r},\tau)f^{\dag}_{\uparrow}(\vec{0},0)\rangle \nonumber \\
		&+& \langle
b^{\dag}_{2}(\vec{r},\tau)b_{2}(\vec{0},0)f^{\dag}_{\downarrow}(\vec{r},\tau)f_{\downarrow}(\vec{0},0)\rangle \Big)
\end{eqnarray*}
The angle brackets $\langle \ldots \rangle$ mean averaging with respect to the
partition function (\ref{SU2_partition}). The complication with this task
however is the presence of light bosons whose tendency to condense prohibits
sensible perturbation theory.  Our strategy here will be to first calculate
the following mean field decomposition 
\begin{eqnarray*}
G(\vec{r},\tau)_{MF} = &-&\frac{1}{2}\langle
b_{1}(\vec{0},0)b^{\dag}_{1}(\vec{r},\tau)\rangle \langle f_{\uparrow}(\vec{r},\tau)f^{\dag}_{\uparrow}(\vec{0},0)\rangle \nonumber \\
		&-& \frac{1}{2}\langle
b_{2}(\vec{0},0)b^{\dag}_{2}(\vec{r},\tau)\rangle \langle f^{\dag}_{\downarrow}(\vec{r},\tau)f_{\downarrow}(\vec{0},0)\rangle 
\end{eqnarray*}
and then incorporate the confining effect of the gauge field by making the
algebraic decay of the spinon Green's function in real space slower. The
motivation for this is rooted in the result of the previous section (the
negative anomalous dimension for the spinon amplitude at the ASL fixed point)
and we will discuss the implied approximations down below.   With the
help of 

\begin{eqnarray*}
G_{f\uparrow}(\vec{k},\omega_n) &=& \sum_{\vec{q'},\nu'_n}\langle f_{\uparrow}(\vec{k},\omega_n)f^{\dag}_{\uparrow}(\vec{q'},\nu'_n)\rangle  \\
G_{f\downarrow}(\vec{k},\omega_n) &=& \sum_{\vec{q'},\nu'_n}\langle f_{\downarrow}(\vec{q'},\nu'_n)f^{\dag}_{\downarrow}(-\vec{k},-\omega_n)\rangle \\
G_{b_1}(\vec{k},\omega_m) &=& \sum_{\vec{q'},\nu'_m}\langle b_{1}(\vec{q'},\nu'_m)b^{\dag}_{1}(-\vec{k},-\omega_m)\rangle \\
G_{b_2}(\vec{k},\omega_m) &=& \sum_{\vec{q'},\nu'_m}\langle b_{2}(\vec{q'},\nu'_m)b^{\dag}_{2}(-\vec{k},-\omega_m)\rangle 
\end{eqnarray*}
we can express the Fourier transform of $G(\vec{r},\tau)_{MF}$ in the form
\begin{eqnarray*}
G(\vec{k},\omega_n) = &-&\frac{1}{2}\int \frac{d\vec{q}d\nu_n}{(2\pi)^3}G_{f\uparrow}(\vec{q},\nu_n)G_{b_1}(\vec{k}-\vec{q},\omega_n-\nu_n) \\
 &+& \frac{1}{2}\int \frac{d\vec{q}d\nu_n}{(2\pi)^3}G_{f\downarrow}(\vec{q},\nu_n)G_{b_2}(\vec{k}-\vec{q},\omega_n-\nu_n)  
\end{eqnarray*}
To make contact with our previous discussions let us concentrate on the
fermionic part of the theory for the moment. From the form (\ref{sflux}) of
the mean field gauge we see that translational invariance is explicitly
broken. After Fourier transformation the fermionic Lagrangian is given in the
$f$ basis by
\begin{eqnarray*}
\label{fLag}
L_{Mf} = {\sum_{\vec{q},\omega_n,\sigma}}^\prime 
\left( f^{\dag}_{\sigma}(\vec{q}),f^{\dag}_{\sigma}(\vec{q}+\vec{Q})\right) 
\quad \quad \quad \quad \quad \quad \\
\left[-i\omega_n\openone + \epsilon_{f}(\vec{q})\sigma_{3} 
+ \eta_{f}(\vec{q})\sigma_{2}\right]{f_{\sigma}(\vec{q}) 
\choose f_{\sigma}(\vec{q}+\vec{Q})}
\end{eqnarray*}
where $\sum^{'}$ denotes a summation over the magnetic Brillouin zone which is
half the size of the original one and takes account of the fact that in the sF
phase the real space unit cell has been doubled with principal axis along
$\pm\hat{x}\pm\hat{y}$.  The $\sigma$ matrices operate in
$\vec{q},\vec{q}+\vec{Q}$ space with $\vec{Q} = (\pi,\pi)$.  Furthermore note
that
\begin{eqnarray}\label{energies}
\epsilon_{f}(\vec{q}) = -2\tilde{J}\chi(cos(q_{x}a)+cos(q_{y}a)) \\ 
\eta_{f}(\vec{q}) = -2\tilde{J}\Delta (cos(q_{x}a)-cos(q_{y}a))
\end{eqnarray}
We can now simply obtain 
\begin{eqnarray}
G_{f\uparrow}(\vec{k},\nu_n) &=& \frac{i\nu_n + \epsilon_f - i\eta_f}{\nu_n^2 + E^2_f} \quad \quad  E^2_f = \epsilon^2_f + \eta^2_f  \\
G_{f\downarrow}(\vec{k},\nu_n) &=& \frac{i\eta_f -i\nu_n + \epsilon_f}{\nu_n^2 + E^2_f} 
\end{eqnarray}
by inverting the matrix in the above Lagrangian $L_{Mf}$. Analogously we find
for the boson Green's functions
\begin{eqnarray*}
G_{b_1}(\vec{k},\omega_m) &=& \frac{i\omega_m - \mu - \epsilon_b - i\eta_b}{(i\omega_m-\mu)^2 - E^2_b} \quad  E^2_b = \epsilon^2_b+\eta^2_b \\
G_{b_2}(\vec{k},\omega_m) &=& \frac{i\omega_m - \mu + \epsilon_b - i\eta_b}{(i\omega_m-\mu)^2 - E^2_b} \\ 
\epsilon_{b}(\vec{q}) &=& -2\tilde{t}\chi(cos(q_{x}a)+cos(q_{y}a)) \\ 
\eta_{b}(\vec{q}) &=& -2\tilde{t}\Delta (cos(q_{x}a)-cos(q_{y}a))
\end{eqnarray*} 
where $\mu$ is the chemical potential. 

 The standard way to proceed is now to calculate the convolution
integrals in momentum space to obtain the mean field Green's function. As
outlined above we will however follow the different route of first calculating
the Fourier transform of the above Green's functions.

 Let us first concentrate on the boson part which can be written as
\begin{eqnarray}\label{Gb1}
G_{b_1}(\vec{k},\omega_m) = \frac{E_b + \epsilon_b + i\eta_b}{2E_b(i\omega_m - \mu + E_b)} + \frac{E_b - \epsilon_b - i\eta_b}{2E_b(i\omega_m - \mu - E_b)}\nonumber \\
\end{eqnarray}
Expanding for small $\vec{k}$ we obtain
\begin{eqnarray*}
E_b(\vec{k}) &\simeq& 4\tilde{t}\chi - \tilde{t}\chi (ka)^2 \\
\epsilon(\vec{k}) &\simeq& -4\tilde{t}\chi + \tilde{t}\chi (ka)^2  \\
\eta_b(\vec{k}) &\simeq& \tilde{t}\Delta a^2[k_x^2 - k_y^2]
\end{eqnarray*}
Since $-\mu-E_b(0) \geq 0$ we can neglect the contribution from the first term
(\ref{Gb1}) and Fourier transform the second to obtain
\begin{eqnarray}\label{Gb1_real}
G_{b_1}(\vec{r},\tau) =  \frac{1}{2}\Theta(-\tau)e^{(\dm)|\tau|} \times \quad \quad \quad \quad   \\ 
\int\frac{d^2k}{(2\pi)^2}e^{i\vec{k}\cdot\vec{r}}\left[1-\frac{\epsilon_b+i\eta_b}{E_b}\right]e^{-\tilde{t}\chi(ka)^2|\tau|} \\
= \Theta(-\tau)e^{(\dm)|\tau|}\frac{m_b}{2\pi|\tau|}e^{-\frac{m_b r^2}{2|\tau|}}
\end{eqnarray}
where $m_b = \frac{1}{2\tilde{t}\chi a^2}$ is the band mass of the boson.
(\ref{Gb1_real}) is nothing but the propagator for a particle of mass $m_b$ in
2d.   Fourier transforming $G_{f\uparrow}(\vec{k},\omega_n)$ with
respect to frequency yields
\begin{equation}
G_{f\uparrow}(\vec{k},\tau) = \frac{1}{2}e^{-E_f|\tau|}\left[ sign(\tau) + \frac{\epsilon_f - i\eta_f}{E_f}\right]
\end{equation}
Next we expand the dispersion about the node \newline $\tilde{{\bf Q_1}} =
(\frac{\pi}{2a},\frac{\pi}{2a})$ 
\begin{eqnarray*}
\epsilon_f(\tilde{{\bf Q}}_1+\tilde{{\bf p}}) &=& 2\chi\tilde{J}a(\tilde{p}_x + \tilde{p}_y) \equiv v_f p_1 \\
\eta_f(\tilde{{\bf Q}}_1+\tilde{\bf p}) &=& -2\Delta\tilde{J}a(-\tilde{p}_x + \tilde{p}_y) \equiv v_2 p_2
\end{eqnarray*}
where $v_f = 2 \sqrt{2}\chi \tilde{J} a \quad v_2 = 2\sqrt{2}\Delta\tilde{J}a$
and we resort to the notation $\vec{p} = {\bf p}$ to denote 2d vectors.

 We now split up the momentum integral $\int
\frac{d^2k}{(2\pi)^2}e^{i\vec{k}\cdot \vec{r}} = \int_{\tilde{\bf
Q}_i}\frac{d^2\tilde{p}}{(2\pi)^2}e^{i(\tilde{\bf Q}_i + \tilde{\bf p})\cdot
\vec{r}}$ where $\tilde{\bf Q}_i$ with $i = 1 ...4$ corresponds to
$(\pi/2,\pi/2), (-\pi/2,-\pi/2), (\pi/2,-\pi/2), (-\pi/2, \pi/2)$ in this
order and we have set $a$ the lattice spacing equal to unity. 

 The momentum Fourier transform about $\tilde{\bf Q}_1$ is then given
by
\begin{eqnarray*}
\int_{\tilde{\bf Q}_1}\frac{d^2\tilde{p}}{(2\pi)^2}e^{i(\tilde{\bf Q}_i + \tilde{\bf p})\cdot \vec{r}} \frac{1}{2} e^{-\sqrt{(v_fp_1)^2 + (v_2p_2)^2}|\tau|}\times\quad \quad \quad \quad \quad\\ \left[ sign(\tau) + \frac{v_fp_1 + iv_2p_2}{\sqrt{(v_fp_1)^2 + (v_2p_2)^2}}\right] = \frac{e^{i \tilde{\bf Q}_1\cdot \vec{r}}}{4\pi v_f v_2}\frac{\left[ \tau + i \tilde{x_1} - \tilde{y_1} \right]}{(\tau^2 + \tilde{\bf r}_1^2)^{\frac{3}{2}}}
\end{eqnarray*}
with $\tilde{x_1} = \frac{x+y}{\sqrt{2}v_f} \quad \tilde{y_1} =
\frac{y-x}{\sqrt{2}v_2} $.

 Analogously we can obtain the contributions around the other three
nodes - hence
\begin{eqnarray*}
G_{f\uparrow}(\vec{r},\tau) = \int\frac{d^2k}{(2\pi)^2}e^{i\vec{k}\cdot\vec{r}}\frac{1}{2}e^{-E_f|\tau|}\left[ sign(\tau) + \frac{\epsilon_f - i\eta_f}{E_f}\right]\\ = \frac{e^{i \tilde{\bf Q}_1\cdot \vec{r}}}{4\pi v_f v_2}\frac{\left[ \tau + i \tilde{x_1} - \tilde{y_1} \right]}{(\tau^2 + \tilde{\bf r}_1^2)^{\frac{3}{2}}} + \frac{e^{i \tilde{\bf Q}_2\cdot \vec{r}}}{4\pi v_f v_2}\frac{\left[ \tau - i \tilde{x_2} + \tilde{y_2} \right]}{(\tau^2 + \tilde{\bf r}_2^2)^{\frac{3}{2}}} \\
+ \frac{e^{i \tilde{\bf Q}_3\cdot \vec{r}}}{4\pi v_f v_2}\frac{\left[ \tau - i \tilde{y_3} + \tilde{x_3} \right]}{(\tau^2 + \tilde{\bf r}_3^2)^{\frac{3}{2}}} + \frac{e^{i \tilde{\bf Q}_4\cdot \vec{r}}}{4\pi v_f v_2}\frac{\left[ \tau + i \tilde{y_4} - \tilde{x_4} \right]}{(\tau^2 + \tilde{\bf r}_4^2)^{\frac{3}{2}}}
\end{eqnarray*}
where $\tilde{x_3} = \tilde{x_4} =  \frac{x+y}{\sqrt{2}v_2} \quad \tilde{y_3}
=\tilde{y_4} =  \frac{y-x}{\sqrt{2}v_f} $ and $\tilde{\bf r}_1 = \tilde{\bf
r}_2$. Identical calculations as given above lead to the expressions for
$G_{f\downarrow}(\vec{r},\tau)$ and $G_{b_2}(\vec{r},\tau)$. Putting all of
them together we finally obtain
\begin{eqnarray}\label{electron_realspace}
G^{e}(\vec{r},\tau) &=& \Theta(-\tau)e^{(\dm)|\tau|}\frac{m_b}{2\pi|\tau|}e^{-\frac{m_b r^2}{2|\tau|}} \times \\ & &\left\{ \frac{e^{i \tilde{\bf Q}_1\cdot \vec{r}}}{4\pi v_f v_2}\frac{\left[ -\tau - i \tilde{x_1}\right]}{(\tau^2 + \tilde{\bf r}_1^2)^{\frac{3}{2}}} + \frac{e^{i \tilde{\bf Q}_2\cdot \vec{r}}}{4\pi v_f v_2}\frac{\left[ -\tau + i \tilde{x_2} \right]}{(\tau^2 + \tilde{\bf r}_2^2)^{\frac{3}{2}}} \right. \nonumber \\
&+& \left. \frac{e^{i \tilde{\bf Q}_3\cdot \vec{r}}}{4\pi v_f v_2}\frac{\left[ -\tau + i \tilde{y_3} \right]}{(\tau^2 + \tilde{\bf r}_3^2)^{\frac{3}{2}}} + \frac{e^{i \tilde{\bf Q}_4\cdot \vec{r}}}{4\pi v_f v_2}\frac{\left[ -\tau - i \tilde{y_4} \right]}{(\tau^2 + \tilde{\bf r}_4^2)^{\frac{3}{2}}} \right\} \nonumber
\end{eqnarray}
At this point let us make the following replacement in the spinon part of the
physical hole propagator.
\begin{equation}
\frac{1}{(\tau^2 + \tilde{\bf r}^2)^{\frac{3}{2}}} \rightarrow \frac{\Gamma^{\alpha}}{(\tau^2 + \tilde{\bf r}^2)^{\frac{3-\alpha}{2}}}
\label{heuristic}
\end{equation}

\begin{figure}[htb]
\begin{center} 
\vfil
\includegraphics[width=70mm]{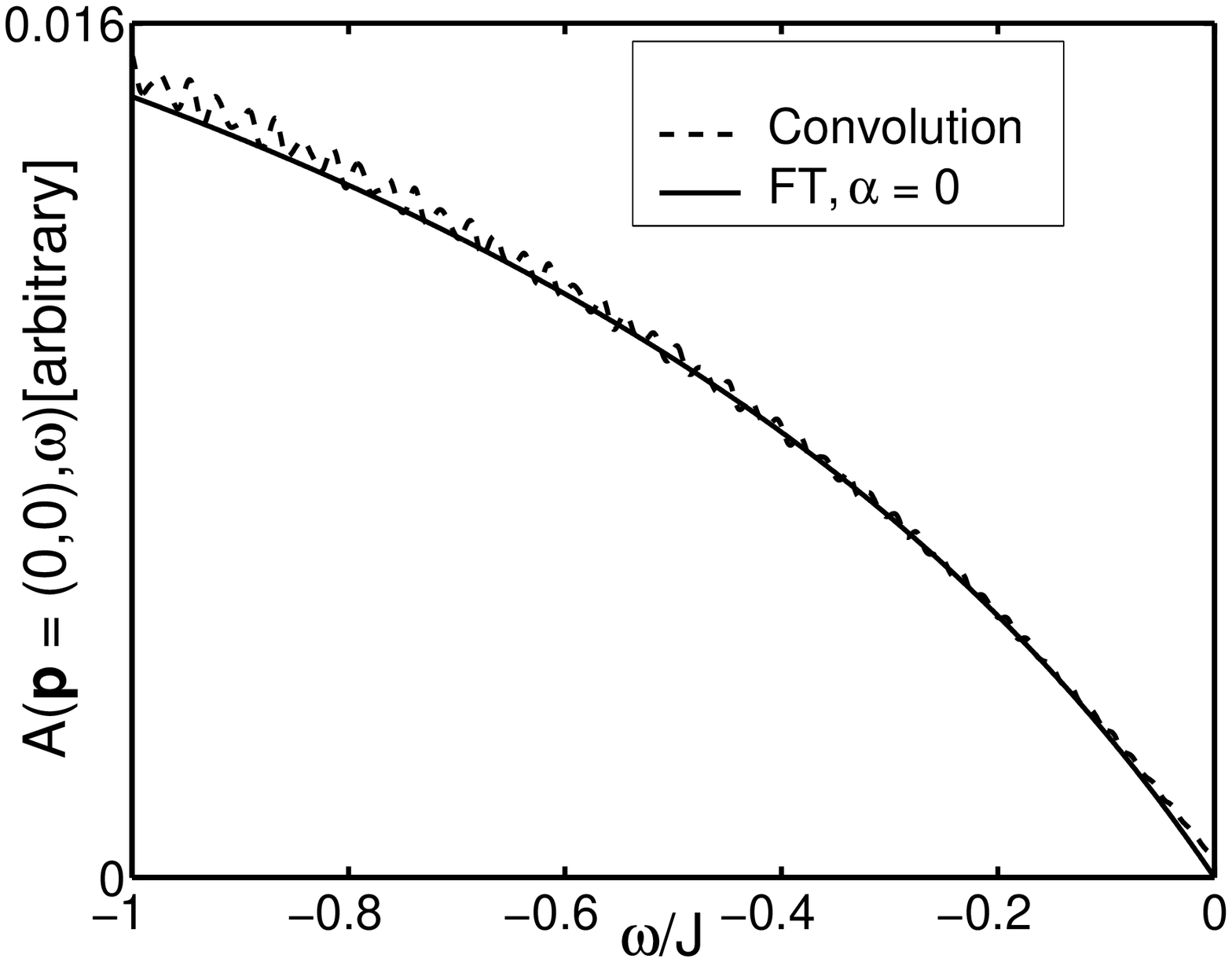}
\vfil
\includegraphics[width=75mm]{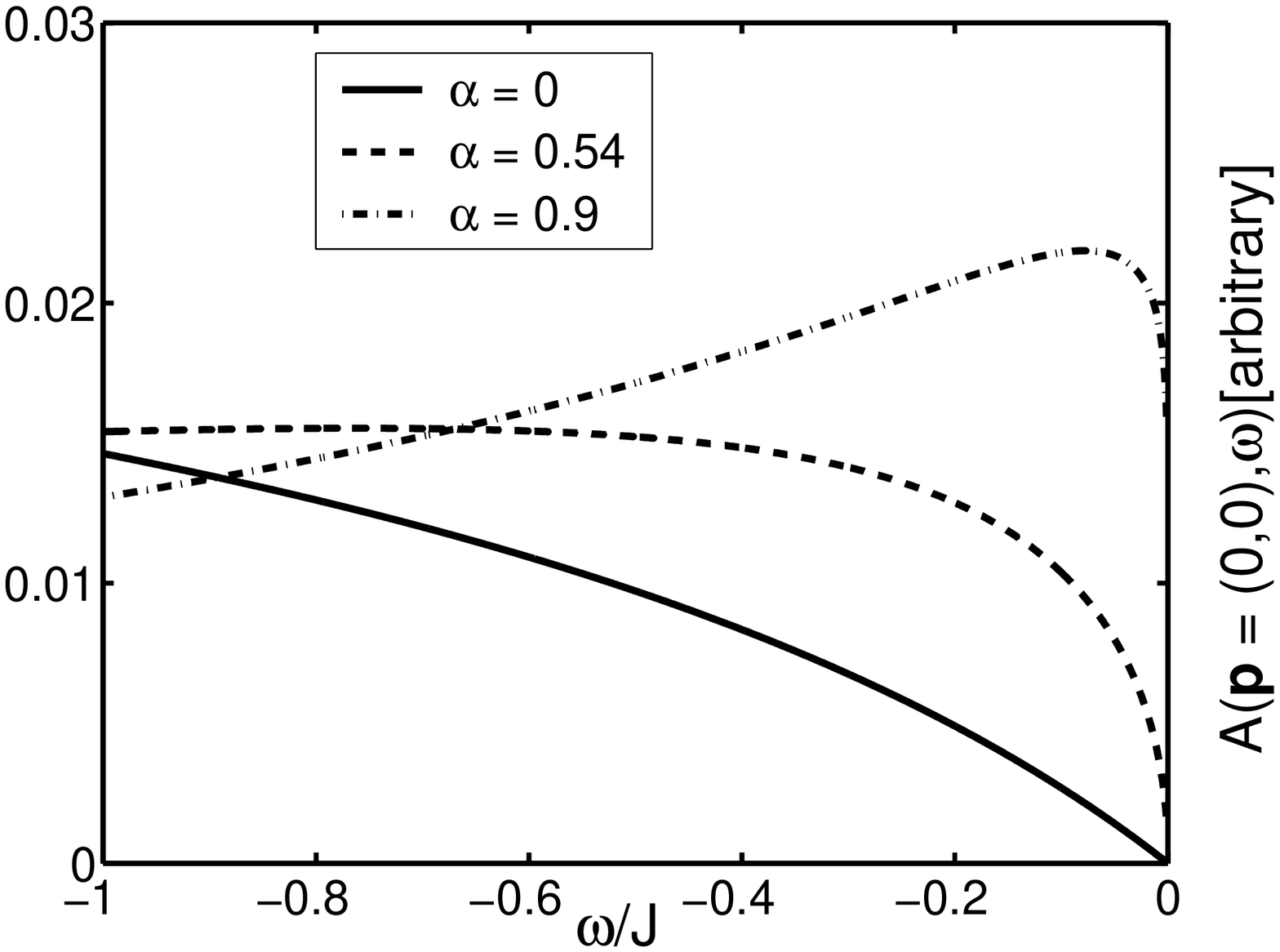}
\vfil
\caption{The top figure shows the good agreement between a direct evaluation
of the convolution integral in momentum space (dashed line) and the result
based on equation (\ref{SP_p_0_mb_arbitrary}) the Fourier Transform with
$\alpha = 0$. The bottom figure depicts the evolution of the spectrum as we
increase $\alpha$. The value $\alpha = 0.54$ is ``motivated'' by our earlier
analysis. Note that here $\tilde{t}/\tilde{J} = 3$ and we have used the
implementation of the hypergeometric function given in ``Numerical Recipes in
C'', W.H. Press \it{et.al} \cite{NR}} 
\label{conv_FT_comparison} 
\end{center}
\end{figure}
\begin{figure}[htb]
\begin{center}
\vfil
\includegraphics[width=75mm]{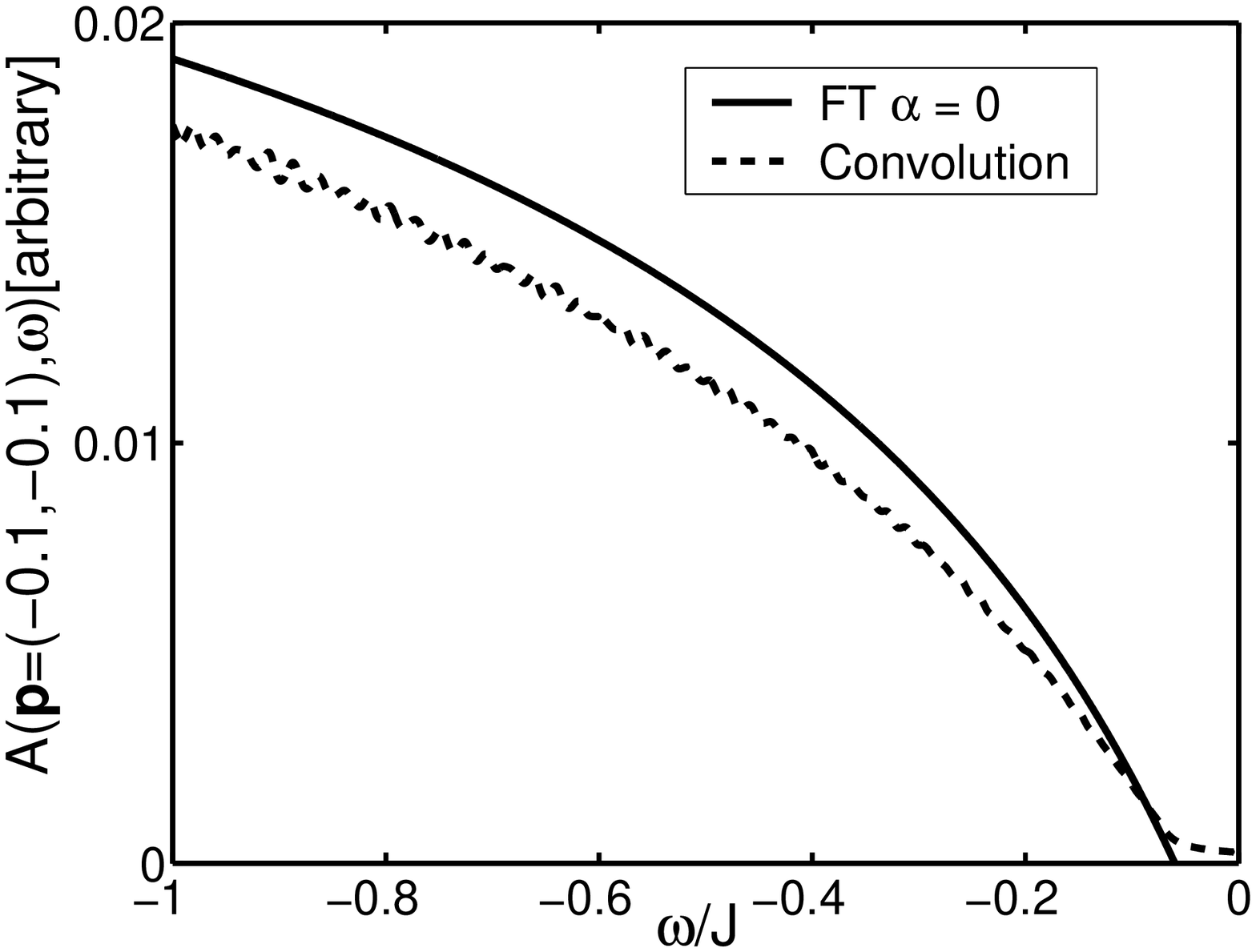}
\vfil
\includegraphics[width=75mm]{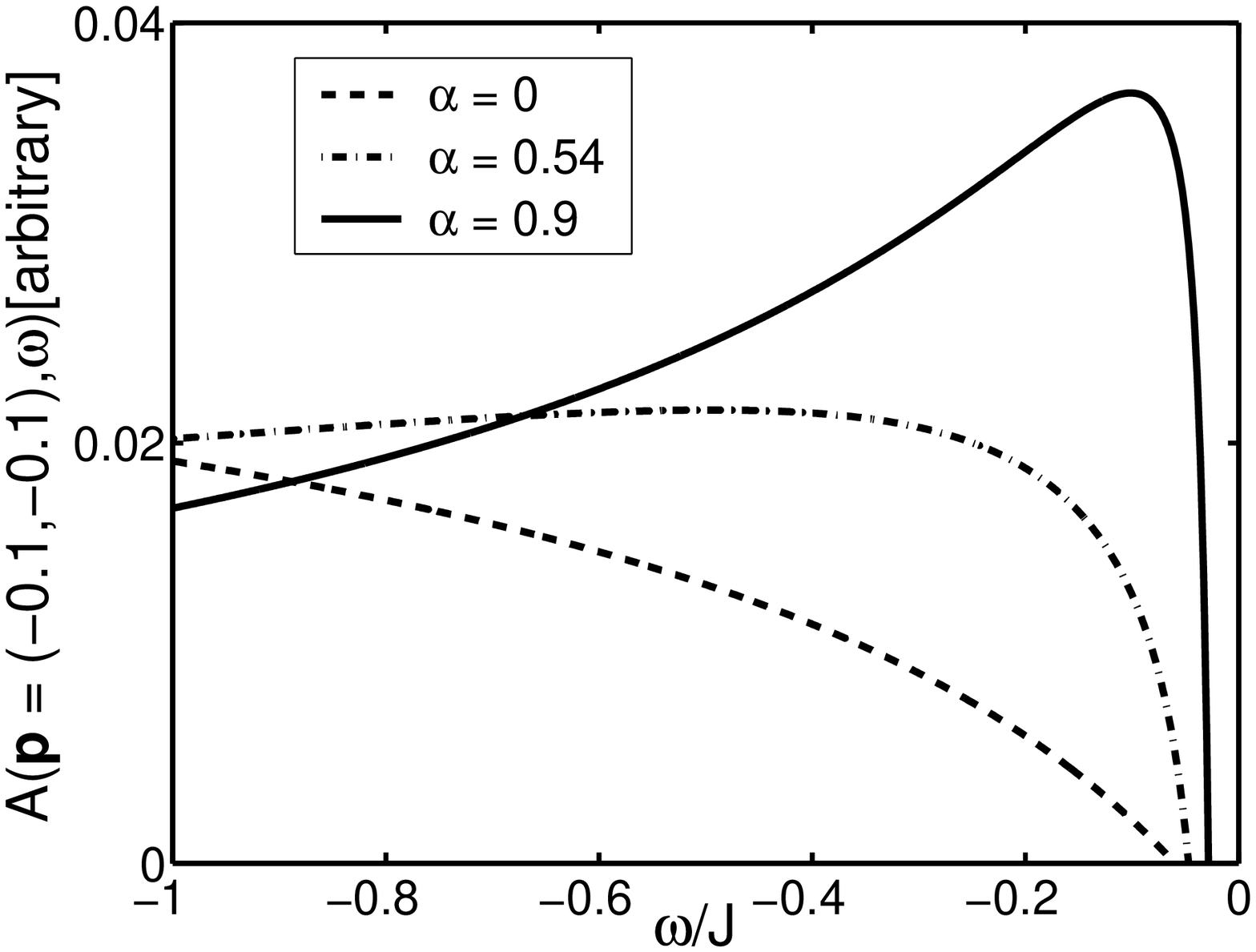}
\vfil
\caption{Top: Direct evaluation of the convolution integral in momentum space
(dashed line) compared with the result based on analytic continuation of
equation (\ref{full_answer}) with $\alpha = 0$. Note that the result of
equation (\ref{full_answer}) actually goes negative for small $\omega$ which
we believe is an artifact of the numerical evaluation  where we sum terms that
alternate in sign but don't form a null sequence. This is supported by the
agreement with the direct evaluation of the convolution integral which is
small as can be seen in the figure in the region for $\omega \rightarrow 0$
where the numerics becomes tricky. The curve is obtained with $n=4$ in the
evaluation of (\ref{full_answer}) which shows the fast convergence for
intermediate values of $\omega$.  The bottom figure depicts the evolution of
the spectrum as we increase $\alpha$. Note that the onset of the spectrum is
determined by $\frac{p^2}{2m_b}$}
\label{Figure5}
\end{center}
\end{figure}

This procedure is motivated by revisiting the following first quantized path
integral formulation for the electron Green's function
\begin{eqnarray*}
 G(\vec{r},\tau) &=&
 N^{-1}
\int D a Dx_fDx_b e^{-i\int_0^{(\vec{r}, \tau)} d\tau (L_f + L_b +L_a)}
\nonumber\\
&& \quad\quad\quad
\quad\quad\quad
e^{i\int_{\wr_b}a\cdot dx-i\int_{\wr_f}a\cdot dx}
\end{eqnarray*}
where $\int_{\wr_b}$ integrates along the boson path $x_b(\tau)$ and
$\int_{\wr_f}$ integrates along the fermion path $x_f(\tau)$.  $L_{f,b,a}$ are
Lagrangians for the fermion, the boson, and the gauge fields.  Let $\mid$ be
the straight line connecting $(0,0)$ and $(\vec{r}, \tau)$.  Since the fermion
is relativistic, the area spanned by the fermion path $\wr_f$ and $\mid$ is of
order $L^2$, where $L^2 = r^2+v_f^2 \tau^2$ and $v_f$ is the fermion velocity.
For the non-relativistic boson, the area spanned by the boson path $\wr_b$ and
$\mid$ is of order $v_f \tau \sqrt{\tau/m_b} $, with $m_b$ the boson mass. For
$\tau > 1/m_b v_f^2$, the area between the loop spanned by $\wr_f$ and $\wr_b$
mainly comes from the area between $\wr_f$ and $\mid$ and we can approximate
$\int_{\wr_b}$ by $\int_\mid$:
\begin{eqnarray}
&& G = N^{-1}\nonumber\\
&&\int Da Dx_fDx_b e^{-i\int_0^{(\vec{r}, \tau)} d\tau (L_f + L_b+L_a)}
e^{i\int_{\mid}a\cdot dx-i\int_{\wr_f}a\cdot dx} \nonumber\\
&& = G_f G_{b0}    \\
&&G_f = \nonumber\\
&&N^{-1}
\int Da Dx_f e^{-i\int_0^{(\vec{r}, \tau)} d\tau (L_f+L_a) }
e^{i\int_{\mid}a\cdot dx-i\int_{\wr_f}a\cdot dx} \nonumber
\label{GGf}
\end{eqnarray}
where $G_{b0}$ is the free boson Green's function (not coupled to the gauge
fields). Note that the fermion Green's function $G_f$ is just the gauge
invariant Green's function discussed in the first part of the paper.  Hence we
see that by performing the above replacement (\ref{heuristic}) with $\alpha =
2\gamma_\Psi \sim 0.54$ we make the transition from the free spinon Green's
function to the particular gauge invariant spinon amplitude discussed in
section \ref{fgreenfunction}. Furthermore we note here that our previous
attempt of identifying the physical hole Green's function with the gauge
invariant spinon amplitude corresponds to neglecting the $\frac{1}{|\tau|}$
term in (\ref{electron_realspace}) which arises from the fluctuations of the
holon about the classical path (the static straight line in the $m_b
\rightarrow \infty$ limit). In what follows we will analyze the full
expression (\ref{electron_realspace}) under the replacement (\ref{heuristic})
where the fluctuations of the holons are assumed decoupled from the gauge
field (hence the {\emph free} boson  $\frac{1}{|\tau|}$ contribution from the
fluctuations around the classical path). This in particular implies that the
area between the loop spanned by $\wr_f$ and $\wr_b$ is solely given by the
relativistic spinon path combined with the straight line segment which from
the previous discussion is only exact in the limit $m_b \rightarrow \infty$ -
the static holon case. Thus the value $\alpha = 2\gamma_\Psi \sim 0.54$
reflects the total gauge fluctuation effects strictly speaking only for the
static holon case.  Granted these limitations of the present analysis let us
press on and analyze the effect of a finite positive $\alpha$. Note also that
we introduced $\Gamma$ with units of energy to balance dimensions whose value
will be on the order of the lattice scale.

 Thus our task now is to calculate
\begin{equation}
G^{e}(\vec{k},\omega_n) = \int d^2r d\tau e^{-i\vec{k}\cdot\vec{r}+i\omega_n\tau}G^{e}(\vec{r},\tau)
\end{equation}
To do this we first concentrate on $\vec{k} = {\bf Q}_1 + \vec{p}$ and note
that the other three points are related by symmetry.  From expression
(\ref{electron_realspace}) we see that only the first term in  curly brackets
contributes since all the other terms have large phases even in the infrared
limit.  We are thus left with the evaluation of
\begin{eqnarray}\label{G_start}
G^{e}({\bf Q}_1+\vec{p},\omega_n) &=& \int d^2r \int_0^\infty d\tau e^{-i \vec{p}\cdot \vec{r}}e^{-[\om]\tau} \nonumber \\
&\times&\frac{m_b L^{-\alpha}}{2\pi\tau}e^{-\frac{m_b r^2}{2\tau}} \frac{\left[ \tau - i \frac{x+y}{\sqrt{2}v}\right]v}{4\pi(\tau^2v^2 + \vec{r}^2)^{\frac{3-\alpha}{2}}} \nonumber \\
\end{eqnarray}
where we have simplified to the case $v_f = v_2 = v$ and have taken care of
the fact that our analysis only allows us to calculate the hole spectral
function which was indicated by the $\Theta(-\tau)$ factor in the boson
propagator.

Next let us analyze expression (\ref{G_start}) in some limiting cases where
the integrals can be performed exactly. The general result and some more
technical points can be found in the appendix.

The simplest point to analyze is the expression right at the node $\vec{p} =
0$ where the terms proportional to $x \& y$ average to zero on angular
integration and we are left with  
\begin{eqnarray*}
G^{e}({\bf Q}_1,\omega_n) &=& K_1 \int_0^\infty dr r \int_0^\infty d\tau e^{-[\om]\tau} \\
&\times& \frac{e^{-\frac{m_b r^2}{2\tau}}}{(\tau^2 v^2 + r^2)^{\frac{3-\alpha}{2}}}
\end{eqnarray*}
where $K_1 = \frac{m_b v L^{-\alpha}}{8 \pi} \quad L = v/\Gamma $. The $r$
integral can be evaluated via \newline Gradshteyn Ryzhik 5 edition (GR5)
\cite{GR} 3.382 (4)
\begin{eqnarray*}
\int_0^\infty dx (x+\beta)^\nu e^{-\mu x} dx = \frac{1}{\mu^{1+\nu}}e^{\beta\mu}\Gamma(\nu+1,\beta\mu) \\
|arg\beta| < \pi, Re \mu > 0 
\end{eqnarray*}
where $\Gamma(x,y)$ is an incomplete Gamma function.  The resulting $\tau$
integral can again be performed with the help of (GR5) 6.455 (1)
\begin{eqnarray*}
\int_0^\infty x^{\mu-1}e^{-\beta x}\Gamma(\nu,\alpha x) dx = \quad \quad \quad \quad \\ \frac{\alpha^\nu \Gamma(\mu+\nu)}{\mu(\alpha+\beta)^{\mu+\nu}} \left. \right._2F_1\left(1,\mu+\nu;\mu+1;\frac{\beta}{\alpha + \beta}\right) \\
Re(\alpha+\beta)>0, Re\mu > 0, Re(\mu+\nu)>0
\end{eqnarray*}
where $\left. \right._2F_1$ is a hypergeometric function and we have the
following identifications $\beta = [\om]-\frac{m_bv^2}{2} \quad \alpha =
\frac{m_b v^2}{2} \quad \mu = \frac{\alpha+1}{2} \quad \nu = \frac{\alpha -
1}{2}$.
Putting this together we arrive at
\begin{eqnarray}
G^{e}({\bf Q}_1,\omega_n) = \frac{m_b \Gamma^{\alpha}}{4\pi} \frac{\Gamma(\alpha)}{(\alpha+1)}\frac{1}{[\om]^\alpha}\\ 
\left. \right._2F_1\left(1,\alpha;\frac{\alpha+3}{2};1-\frac{m_b v^2}{2[\om]}\right) \nonumber
\end{eqnarray}
We are now ready to perform the analytic continuation $i\omega_n \rightarrow
\omega - i0^+$ to obtain the hole spectral function at the node ${\bf Q}_1$

\begin{widetext} 
\begin{eqnarray}\label{SP_p_0_mb_arbitrary}
A_{-}({\bf Q}_1,\omega) &=&\frac{m_b \Gamma^{\alpha}}{4\pi}\frac{\Gamma(\alpha)}{(\alpha+1)}\Theta(-\omega)\left[ \frac{sin(\pi\alpha)}{\pi}[\omcon]^{-\alpha}\Theta(\omcon) \right. \nonumber \\
& &Re \left. \right._2F_1\left(1,\alpha;\frac{\alpha+3}{2};1-\frac{m_b v^2}{2[\omega+|\mu|-4\tilde{t}\chi -i0^{+}]}\right)  \\
&+& \left.\frac{cos(\pi\alpha)}{\pi}[\omcon]^{-\alpha}\Theta(\omcon)\right. \nonumber \\
& &\left. Im \left. \right._2F_1\left(1,\alpha;\frac{\alpha+3}{2};1-\frac{m_b v^2}{2[\omega+|\mu| - 4\tilde{t}\chi -i0^{+}]}\right) \right] \nonumber
\end{eqnarray}
\end{widetext}
Note that in principal we also have a contribution from
$\Theta(|\mu|-4\tilde{t}\chi + \omega)$ however since the chemical potential
is exponentially close to the bottom of the band we will set $|\mu| \sim
4\tilde{t}\chi$ for our calculation. It is nevertheless important in the
derivation of the above that $|\mu| - 4\tilde{t}\chi = 0^{+}$ for the
integrals to converge.  In Fig.[\ref{conv_FT_comparison}] we show the effect
of a non-zero $\alpha$ on the spectrum at the node. The top figure simply
compares the two results for the spectrum based on a direct evaluation of the
convolution between single boson and spinon Green's functions (dashed line)
and the evaluation of Eq.(\ref{SP_p_0_mb_arbitrary}) for the case $\alpha \sim
0$ where they should agree.  The bottom portion of
Fig.[\ref{conv_FT_comparison}] depicts the evolution of the spectrum as we
increase $\alpha$ away from zero. Though still rather incoherent a shift of
spectral weight toward lower energies is apparent. This piling up of low
energy spectrum is directly associated with a nonzero alpha which in turn is a
consequence of the gauge interaction between the holon and the spinon
responsible for the incoherent mean-field spectrum. Thus we see that when our
reference system is the incoherent spectrum associated with the independent
propagation of spinon and holon the gauge fluctuations have the desirable and
intuitive effect of binding the two degrees of freedom back into the physical
hole. This should be compared with our initial attempt (in the first half of
the paper) to reference the gauge fluctuation effects with respect to simply a
free spinon plus a coherent boson. A nearly coherent boson (with a well
defined phase) is however not consistent with a massless gauge field whose
origin lies in the strong local phase fluctuations of {\bf both} holon and
spinon.

Fig.[\ref{Figure5}] shows a similar comparison for momentum $p_x = p_y = -0.1$
along the diagonal. This result is based on the general equation derived in
Appendix C. 
The evolution with $\alpha$ follows the expected trend. The onset of the spectrum is governed by $\frac{p^2}{2m_b}$ which gives rise to the peak in the spectrum centered at the boson energy.

\section{Evolution with $m_b$}

A finite $\alpha$ leads to a piling up of spectrum at the lower edge. However
as we pointed out above this onset is governed by the boson energy since the
spinon spectrum is linear with velocity $v$ the minimum of the spectrum for
small momenta is always going to be determined by the boson with its quadratic
dispersion. This of course reflects our cavalier treatment of boson-boson
correlations which in the absence of a full theory for the spinon holon +
gauge system we have totally neglected. Boson correlations will tend to reduce
the density of low energy boson states which are responsible for the large
distance behavior of the boson correlater. On the crudest level we might hope
to capture this reduction by simply decreasing $m_b$ which determines the free
boson density of states. 

 In order to see this dependence most clearly let us re-derive an expression
for the spectrum at the node. Introducing $\tilde{\tau} = \tau v$ we would
like to evaluate
\begin{eqnarray*}
G^{e}({\bf Q}_1,\omega_n) = \int_0^\infty d\tilde{\tau}\int_0^\infty d^2 r e^{-[\om]\frac{\tilde{t}}{v}} \\
\frac{m_b}{8 \pi^2} e^{-\frac{m_b v r^2}{\tilde{\tau}}}[\tilde{\tau}^2 + r^2]^{\frac{\alpha-3}{2}}
\end{eqnarray*}
We can now view this as a 3d integral over the positive hemisphere where
$\tilde{\tau} = R cos\theta \quad r = R sin\theta$ and after integrating out
$\phi$ and $R$ we are left with
\begin{eqnarray*}
G^{e}({\bf Q}_1,\omega_n) = \quad \quad \quad \quad \quad \quad \\
 \frac{K_1}{v} \int_0^1 dx \frac{\Gamma(\alpha)}{[\frac{x}{v}(\om) + \frac{m_b v}{2x}(1-x^2)]^{\alpha}}
\end{eqnarray*}
After analytic continuation we arrive at the following expression for the hole
spectral function 
\begin{widetext}
\begin{eqnarray*}
 A_{-}({\bf Q}_1,\omega) = \Theta(-\omega)\frac{K_1}{v} \frac{sin\pi\alpha \Gamma(\alpha)}{\pi} \int_0^1 dx \Theta\left[|\omega| - |\mu| + 4\tilde{t}\chi + \frac{m_b v^2}{2} - \frac{m_b v^2}{2 x^2}\right] 
 \frac{x^\alpha v^\alpha}{[x^2(\omcon) - \frac{m_b v^2}{2}(1-x^2)]^{\alpha}}
\end{eqnarray*} 
\end{widetext}
which in the $\alpha \rightarrow 0^+$ limit reduces to
\begin{eqnarray}\label{linear}
 A_{-}({\bf Q}_1,\omega) &=& \Theta(-\omega)\frac{K_1}{v} \int_{\sqrt{\frac{m_b v^2}{2[\omcon + m_b v^2/2]}}}^1 dx \nonumber \\
\lim_{\omega \rightarrow |\mu|-4\tilde{t}\chi}& &\Theta(-\omega)\frac{K_1}{v} \frac{|\omega|-|\mu| + 4\tilde{t}\chi}{m_b v^2}
\end{eqnarray}
We thus find that $\frac{1}{m_b}$ determines the slope of the spectrum at the
onset.  In Fig.[\ref{Masstozero}] we depict the effect of decreasing the mass
of the exponential by a factor of 3 ($t = 9J$) and 4 respectively for the case
$\alpha = 0$. Note that in addition to becoming steeper in accord with the
result just derived for the case $\vec{p} = 0$ the onset of the spectrum is
pushed to higher frequencies initially simply following the increase in
$\frac{p^2}{2m_b}$. The curves seem to converge to a step function, however,
the onset of the spectrum does not converge to $p v$ the energy of the spinon
but rather $\frac{pv}{2}$. 

\begin{figure}[htb]
\begin{center} 
\includegraphics[width=80mm]{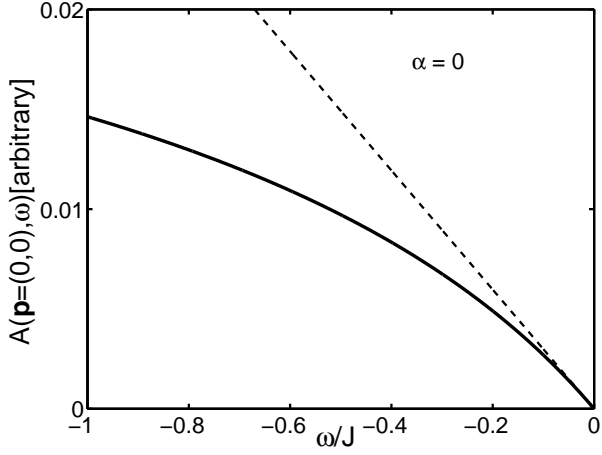}
\caption{Linear approximation Eq.(\ref{linear}) compared with full result for the case $\alpha = 0$.}
\label{Linear_approx}
\end{center}
\end{figure}


\section{$m_b = 0$}

As was indicated in the previous section the free boson dispersion determines
both the onset an the shape of the low energy spectrum. Correlations will
however reduce the low energy modes whose effect we tried to emulate by
reducing the boson density of states. Let us now set $m_b = 0$ from the start.
The rational behind this section is the attempt to capture the correlations
corresponding to collective bosons that are not condensed. For nearly
condensed bosons the appropriate collective variables are the density and
phase and we shall assume that we can neglect fluctuations in the density and
only the correlations of the phase field determine the infrared physics.
\begin{displaymath}
\langle b^{\dagger}(\tau,\vec{r})b(0,0)\rangle = \rho_0\langle e^{-i\theta(\tau,\vec{r})}e^{i\theta(0,0)}\rangle
\end{displaymath}
where the dynamics of $\theta$ is given  by a $2d$ $XY$-model. The $\theta$
Green's function can thus be determined and the resulting boson correlation
function reads
\begin{equation}
\langle b^{\dagger}(\tau,\vec{r})b(0,0)\rangle = \rho_0 e^{-\frac{1}{4\pi v_b\chi_b l}} e^{\frac{1}{4\pi v_b\chi_b\sqrt{r^2+v_b^2\tau^2}}} 
\end{equation}
which in the infrared limit we can write suggestively as
\begin{eqnarray}
\langle b^{\dagger}(\tau,\vec{r})b(0,0)\rangle \simeq \rho_0 e^{-\frac{1}{4\pi v_b\chi_b l}}\left( 1 + \frac{1}{4\pi v_b\chi_b\sqrt{r^2+v_b^2\tau^2}}\right)\nonumber \\
\end{eqnarray}
We will take the second term in parenthesis as reflecting the uncondensed but
correlated boson contribution to the long distance behavior of the boson
propagator. $\chi_b$ denotes the boson compressibility $v_b$ is the velocity
of sound of the collective bosons and $l$ is some lattice scale cut-off
determining the wavefunction renormalization  which will give rise to a
reduction of the superfluid density $\rho_0$.

\begin{figure}[htb]
\begin{center} 
\includegraphics[width=80mm]{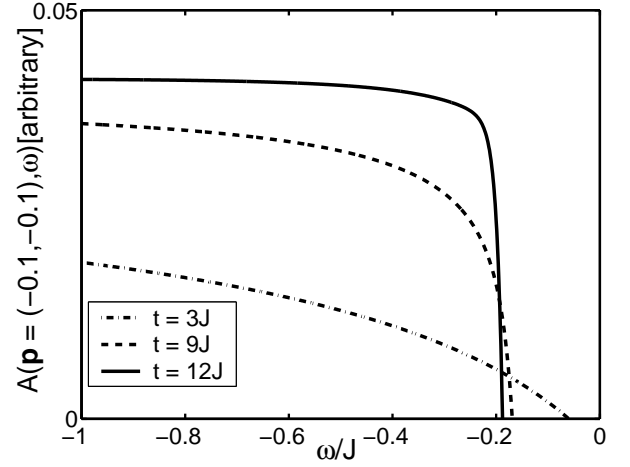}
\caption{Evolution of spectrum with $m_b$ at ${\bf p}= (-0.1,-0.1)$ at $\alpha = 0$}
\label{Masstozero}  
\end{center}
\end{figure}

Taking the above correlation function as the holon part of the physical hole
Green's function we are faced with the evaluation of\begin{widetext}
\begin{eqnarray}\label{boson_linear}
G^{e}({\bf Q}_1+\vec{p},\omega_n) = L^{-\alpha}\int d^2r \int_0^\infty d\tau e^{-i \vec{p}\cdot \vec{r}}e^{-[\om]\tau}\rho_0 e^{-\frac{1}{4\pi v_b\chi_b l}}
\frac{1}{4\pi v_b\chi_b\sqrt{r^2+v_b^2\tau^2}}\frac{\left[ \tau - i \frac{x+y}{\sqrt{2}v}\right]v}{4\pi(\tau^2v^2 + \vec{r}^2)^{\frac{3-\alpha}{2}}} \nonumber
\end{eqnarray}
\end{widetext}
Let us first use the exponential parametrization 
\begin{displaymath}
\frac{1}{(r^2+v^2\tau^2)^\mu} = \frac{1}{\Gamma(\mu)}\int_0^{\infty} ds s^{\mu-1}e^{-s(r^2+v^2\tau^2)}
\end{displaymath}
which allows us to perform both the r integration via (GR5) 6.631 (4) 
\begin{displaymath}
\int_0^\infty dr r^{\nu+1}J_{\nu}(\beta r) e^{-\alpha r^2} = \frac{\beta^\nu}{(2\alpha)^{\nu+1}} exp\left(-\frac{\beta^2}{4\alpha}\right)
\end{displaymath}
and the $\tau$ integral via (GR5) 3.462 (1)
\begin{eqnarray*}
\int_0^\infty x^{\nu-1}e^{-\beta x^2 - \gamma x}dx &=& (2\beta)^{-\nu/2}\Gamma(\nu)exp\left(\frac{\gamma^2}{8\beta}\right)\\
&\times&D_{-\nu}\left(\frac{\gamma}{\sqrt{2\beta}}\right)
\end{eqnarray*}
where $D_\nu$ is a parabolic cylinder function.

\begin{figure}[tb]
\begin{center} 
\includegraphics[width = 80mm]{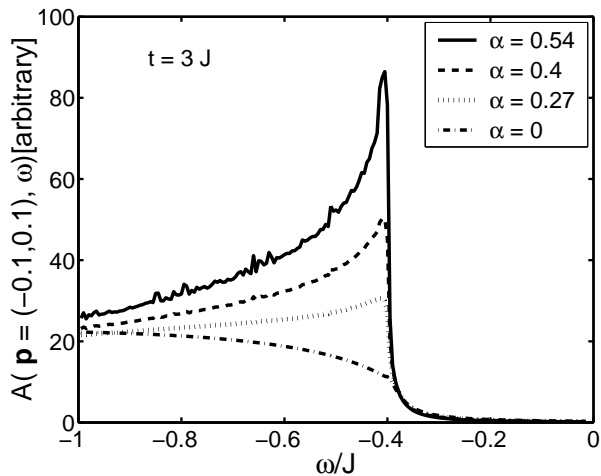}
\caption{Evolution of spectrum with $\alpha$ at $p_x = -p_y = -0.1$. In order to get reasonably smooth graphs we first calculated the spectra with a resolution $\Delta \omega = 0.001J$ and then averaged 5 points to plot the spectrum respectively.} 
\label{gapdirection}
\end{center}
\end{figure}

\begin{figure}[tb]
\begin{center}
\includegraphics[width = 80mm]{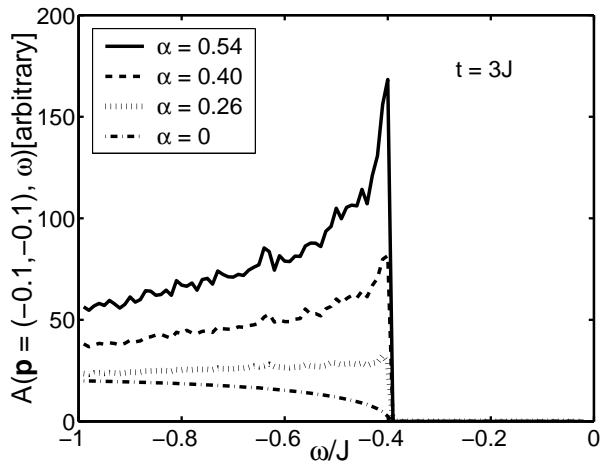}
\caption{Evolution of spectrum with $\alpha$ at $p_x = p_y = -0.1$ along the diagonal. We averaged 10 data points to plot the spectrum which corresponds to $\Delta \omega = 0.01J$}
\label{nodaldirection}
\end{center}
\end{figure}

Once the $r$ and $\tau$ integrals are done we are left with the integrals over
the parameters $s,u$. Concentrating for the moment on the term proportional to
$\tau$ in the numerator we obtain after $r$ and $\tau$ integration
\begin{eqnarray*}
\frac{2\pi K}{\Gamma\left(\frac{3-\alpha}{2}\right)\Gamma\left(\frac{1}{2}\right)} \int_0^\infty ds du s^{-1/2}u^{\frac{1-\alpha}{2}} \quad \quad \quad \times\\
\frac{e^{-\frac{p^2}{4(s+u)}}e^{\frac{[\om]^2}{8(sv_b^2+uv^2)}}}{4(s+u)(sv_b^2+uv^2)} D_{-2}\left(\frac{[\om]}{\sqrt{2(sv_b^2+uv^2)}}\right)
\end{eqnarray*}
Next we perform a change of variables
\begin{displaymath}
v_b^2s = t_1s_1 \quad \quad v^2u = (1-t_1)s_1
\end{displaymath}
which changes the parameter integration to
\begin{eqnarray*}
\frac{2\pi K}{\Gamma\left(\frac{3-\alpha}{2}\right)\Gamma\left(\frac{1}{2}\right)}\int_0^1 dt_1 \int_0^\infty ds_1 \frac{s_1}{v^2v_b^2}\left(\frac{t_1s_1}{v_b^2}\right)^{-1/2}\\
\times \left(\frac{(1-t_1)s_1}{v^2}\right)^{\frac{1-\alpha}{2}}\frac{e^{-\frac{p^2}{4(t_1/v_b^2 + (1-t_1)/v^2)s_1}}}{4s_1^2\left(\frac{t_1}{v_b^2}+ \frac{1-t_1}{v^2}\right)} \quad \quad \quad \\
\times exp\left(\frac{[\om]^2}{8s_1}\right) D_{-2}\left(\frac{[\om]}{\sqrt{2s_1}}\right)
\end{eqnarray*}
Finally we can perform the integral over $s_1$ via (GR) 7.725 (6)
\begin{eqnarray*}
& &\int_0^\infty e^{-zt}t^{-1+\frac{\beta}{2}}D_{-\nu}[2(kt)^{\frac{1}{2}}]dt = \\ & &\frac{2^{1-\beta-\frac{\nu}{2}}\sqrt{\pi}\Gamma(\beta)}{\Gamma\left(\frac{\nu}{2} + \frac{\beta}{2} + \frac{1}{2}\right)}(z+k)^{-\frac{\beta}{2}}\left.\right._2F_1\left(\frac{\nu}{2},\frac{\beta}{2};\frac{\nu+\beta+1}{2};\frac{z-k}{z+k}\right)
\end{eqnarray*}
with the identification $\beta = \alpha \quad \nu = 2 \quad k =
\frac{[\om]^2}{8} \quad z = \frac{p^2}{4\left(\frac{t_1}{v_b^2} +
\frac{1-t_1}{v^2}\right)} - \frac{[\om]^2}{8}$. It turns out that for the
numerical evaluation of the final parameter integral over $t_1$ it is easier
if we use the following quadratic transformation formula for the
hypergeometric function
\begin{eqnarray*}
F(a,b;a-b+1;z) = (1-z)^{-a} \times \quad \quad \\F\left(\frac{a}{2},\frac{a}{2}-b+\frac{1}{2};a-b+1;-\frac{4z}{(1-z)^2}\right)
\end{eqnarray*}
Putting everything together we obtain for the hole Green's function
\begin{widetext}
\begin{eqnarray}
& &G^{e}({\bf Q}_1+\vec{p},\omega_n) = K \left\{ \right.\\
& &\left. \int_0^1 dt t^{-\frac{1}{2}}(1-t)^{\frac{1-\alpha}{2}}\left(\tbar\right)^{\frac{\alpha-2}{2}}\frac{1}{\left([\om]\sqrt{\tbar} + p\right)^{\alpha}} \right. \nonumber \\
& &\left. \left.\right._2 F_1\left(\alpha,\frac{\alpha-1}{2};\frac{3+\alpha}{2}; \frac{[\om]\sqrt{\tbar}-p}{[\om]\sqrt{\tbar}+p}\right) - \frac{\alpha v(p_x+p_y)\sqrt{2}}{v^2} \times \right. \nonumber \\
& &\left.\int_0^1 dt t^{-\frac{1}{2}}(1-t)^{\frac{1-\alpha}{2}}\left(\tbar\right)^{\frac{\alpha-3}{2}}\frac{1}{\left([\om]\sqrt{\tbar} + p\right)^{1+\alpha}}  \right. \nonumber \\
& &\left. \left.\right._2 F_1\left(1+\alpha,\frac{\alpha+1}{2};\frac{3+\alpha}{2}; \frac{[\om]\sqrt{\tbar}-p}{[\om]\sqrt{\tbar}+p}\right)\right\} \nonumber
\end{eqnarray}
\end{widetext}
and $K = \frac{\Lambda^{\alpha}\rho_0e^{-\frac{1}{4\pi v_b\chi_bl}}\Gamma(\alpha)}{2^{2-\alpha}8\pi\chi_bv^2v_b^2\Gamma(\frac{3+\alpha}{2})\Gamma(\frac{3-\alpha}{2})}$. 
We can now continue $i\omega_n \rightarrow \omega -i0^{+}$ analytically and
take the imaginary part to obtain the spectral function. In
Fig.[\ref{gapdirection}] and Fig.[\ref{nodaldirection}] we plot the spectra
for two momenta at $p_x = -p_y = -0.1$ and $p_x = p_y = -0.1$ at different
values of $\alpha$. As opposed to the spectra with finite boson mass
Fig.[\ref{gapdirection}] and Fig.[\ref{nodaldirection}] show the spectral
onset at the spinon energy $v \sqrt{p_x^2+p_y^2}$. We can easily understand
this by noting that $v_b^2 = \frac{\mu}{m_b}$ which with the mean-field values
for $\mu$ and $m_b$ is equal to $\left(\frac{\tilde{t}}{\tilde{J}}\right)^2 v^2$. This in
turn implies that for a given momentum the onset in the spectrum occurs when
all the momentum is taken up by the spinon resulting in a lower edge of the
spectrum tracing out the spinon dispersion \cite{Laughlin97}. As with the case
of finite boson mass the spectra for zero $\alpha$ are broad without any
special features at the onset of the spectrum. The confining tendency of the
gauge fluctuations can again be nicely observed on increasing $\alpha$.      

\section{Conclusion}

The topic of this paper - the single particle spectral function - was analyzed
in a myriad of ways. After briefly addressing the problem of gauge invariance
in the context of the single particle Green's function we gave a first
quantized analysis of a particle coupled to gauge fluctuations. The proposed
gauge invariant quantity to reflect the propagation of the particle was a
loop, one segment of which was made up by the particle path propagating from
$\vec{r}_i$ to $\vec{r}_f$. The loop was completed by a straight line segment
connecting the two destinations. We pointed out that if a particular gauge (an
axial gauge with $\vec{a}\cdot\vec{n} = 0$ with $\vec{n} \propto \vec{r}_f -
\vec{r}_i$) was chosen for the gauge propagator the contribution of the
straight line segment to the particle amplitude could be made to vanish. With
this idea we then analyzed the amplitude in second quantized formulation and
obtained the dressed Green's function with a slower decay in space-time.
However this approach has some spurious infrared problems related to the
incomplete gauge fixing in the temporal gauge and required special (sometimes
ad hoc) regularizations (Leibbrandt prescription). To confirm the result, we
then resorted to discuss the above amplitude as a gauge invariant two-body
propagator with one ``body'' infinitely heavy and hence giving rise to the
straight line path introduced by hand earlier. Thus freed from choosing a
particular gauge we performed a proper analysis of this propagator. The
resulting anomalous dimension for the gauge invariant amplitude agrees with
the previous calculation. It however could not and should not be interpreted
as representing the propagation of a physical particle.

 This forced on us a reinterpretation of the gauge field - not primarily as a
scattering mechanism - but rather a confining field which wants to form a
bound state between the two charged particles propagating under the mutual
attraction mediated by it.  This change in perspective led to the real space
analysis of the second part where we incorporated the gauge fluctuation effect
heuristically into a slower decay of the spinon contribution to the hole
propagator. We saw how this change in the spinon Green's function led to the
piling up of low energy spectral weight. We associate this with the confining
tendencies of the gauge field.  Our treatment of the bosonic holons as free,
however clearly overestimated the density of low energy holon states which led
to the spectrum to be peaked about the holon energy $\frac{p^2}{2 m_b}$. This
is not seen in experiments ( with the possible exception of the single
hole spectrum which seems to have a parabolic band near ($\pi/2,\pi/2$)
\cite{Wells_chapter3} compared with the Dirac spectrum observed at higher
doping levels). In order to overcome this we argued for the holon Green's
function to be determined in the infrared limit by collective correlations
rather than the free particle behavior.  This remedied the problem of the edge
of the spectrum tracing out $\frac{p^2}{2m_b}$ and led to the spectra depicted
in Fig.[\ref{gapdirection}] and Fig.[\ref{nodaldirection}]. It is however
clear that at the level of our present study the real value of  $\alpha$ is
not known and hence we are forced to treat it as a phenomenological parameter.
A proper microscopic analysis of the holon + spinon + gauge system is needed
to put the heuristic results proposed in this paper on a firm foundation. 

We would like to remark that the electron spectral function discussed here is
the high energy spectral function, which is valid for high energies and/or
high temperatures. This is because we have ignored the boson condensation and
the instanton effect. For $T<150K$, the boson-condensation/instanton-effect
will be important which will lead to a spin-charge recombination\cite{CHI} due
to the confinement of $U(1)$ gauge field. In this case a $\delta$-function
peak will appear in the electron spectral function, and low energy excitations
will be described by electron-like quasiparticles.  However, the Fermi
surfaces of those well defined quasiparticles are formed by four small
segments.\cite{WenLee}

We would like to thank P.A. Lee, T. Senthil and A. Seidel for helpful discussions. This research is supported by NSF Grant No. DMR--01--23156 and by NSF-MRSEC Grant No. DMR--98--08941

\begin{widetext}

\section{Appendix A: Fermion self energy in the $a_{0} = 0$ gauge}\label{AppA}

In this appendix we present details of the evaluation of the self energy in
the $a_{0} = 0$ gauge at T=0

\begin{eqnarray}
\frac{N}{8}i\Sigma(\vec{p}) = \gamma^{0}\int \frac{d^{3}q}{(2\pi)^{3}}\Big[
\frac{(q_{0}-p_{0}){\bf q}^{2} + 2 (q_{0}-p_{0})
q_{0}^{2}}{\sqrt{\vec{q}^{2}}(\vec{p}-\vec{q})^{2}q_{0}^{2}}\Big]\nonumber
\\
+\gamma^{x}\int \frac{d^{3}q}{(2\pi)^{3}}\Big[
\frac{(p_{x}-q_{x})(q_{x}^{2}-q_{y}^{2}) +
2q_{x}q_{y}(p_{y}-q_{y})}{\sqrt{\vec{q}^{2}}
(\vec{p}-\vec{q})^{2}q_{0}^{2}}\Big]\nonumber \\
+\gamma^{y}\int \frac{d^{3}q}{(2\pi)^{3}}\Big[
\frac{(p_{y}-q_{y})(q_{y}^{2}-q_{x}^{2}) +
2q_{x}q_{y}(p_{x}-q_{x})}{\sqrt{\vec{q}^{2}}
(\vec{p}-\vec{q})^{2}q_{0}^{2}}\Big]
\end{eqnarray}

Let us take a closer look at the $\gamma^{0}$-term which has two contributions
\begin{equation}
\int \frac{d^{3}q}{(2\pi)^{3}}\left[
\underbrace{\frac{(q_{0}-p_{0})\vec{q}^{2}}{\sqrt{\vec{q}^{2}}
(\vec{p}-\vec{q})^{2}q_{0}^{2}}}_{\mathbb{A}} + 
\underbrace{\frac{(q_{0}-p_{0})}{\sqrt{\vec{q}^{2}}
(\vec{p}-\vec{q})^{2}}}_{\mathbb{B}}\right]
\end{equation}
First look at term $\mathbb{B}$ which can be calculated in the standard way
using dimensional regularization combined with the Feynman trick in the
generalized form
\begin{displaymath}
\frac{1}{A^{\alpha}B^{\beta}} = \int_{0}^{1}dz
\frac{z^{\alpha-1}(1-z)^{\beta-1}}{[zA+(1-z)B]^{\alpha + \beta}}
\frac{\Gamma(\alpha +\beta)}{\Gamma(\alpha)\Gamma(\beta)}
\end{displaymath}
With $ A = (\vec{p}-\vec{q})^{2} \quad B = \vec{q}^{2}$ we find for
$\mathbb{B}$
\begin{displaymath}
\int \frac{d^{3}q}{(2\pi)^{3}}\frac{ 
2 (q_{0}-p_{0})}{\sqrt{\vec{q}^{2}}(\vec{p}-\vec{q})^{2}}|_{div} 
= - \frac{p_{0}}{6 \pi^2}\frac{1}{3-d}
\end{displaymath}

Term $\mathbb{A}$ exhibits in addition to the UV divergence which presumably
will be taken care of by some appropriate regularization procedure an infrared
divergence due to the $q_{0}^2$ in the denominator. This divergence is an
indication that our gauge fixing $a_{0}=0$ has a large residual gauge freedom
in the form of time independent gauge transformations. This reflects the fact
that in using the temporal gauge we have ``lost'' Gauss' law which within the
Lagrangian formulation is obtained as a constrained on the gauge fields by
varying the action with respect to $a_{0}$.  In order to deal with this
problem we follow Leibbrandt and Staley \cite{LS_chapter3} which have
introduced a prescription for this unphysical pole in the form
\begin{displaymath}
\frac{1}{q_{0}^2} \rightarrow lim_{\mu\rightarrow 0} \frac{q_0}{(q_{0}^2 + \mu^2)^2}
\end{displaymath}
where the limit $\mu \rightarrow 0$ is taken after the integrations over loop
momenta and Feynman parameters is completed.  This leaves us with the
following expression for $\mathbb{A}$
\begin{equation}
\mathbb{A} = lim_{\mu\rightarrow 0} \int \frac{d^{3}q}{(2\pi)^{3}}
\frac{(q_{0}-p_{0})\vec{q}^{2}}{\sqrt{\vec{q}^{2}}
(\vec{p}-\vec{q})^{2}}\frac{q_{0}^2}{(q_{0}^2 + \mu^2)^2}
\end{equation}

To compute this integral \cite{Leibbrandt87} we first combine the denominator
using exponential parametrization
\begin{displaymath}
\frac{1}{(\vec{p}-\vec{q})^{2}\sqrt{\vec{q}^{2}}(q_{0}^2 + \mu^2)^2} = \int_{0}^{\infty} dt_1 dt_2 dt_3 \frac{t_3}{\sqrt{\pi t_2}} exp\left[ -t_1 (\vec{p}-\vec{q})^{2} -t_2 \vec{q}^2 - t_{3}(q_0^2 + \mu^2)\right]
\end{displaymath}  
then we use dimensional regularization to  compute
\begin{eqnarray*}
& &\int \frac{d^d q}{(2\pi)^d} \left[\frac{d}{d(t_1+t_2)}\frac{d}{dt_3}\right] e^{\left[-(t_1+t_2)\vec{q}^2 + 2t_1 \vec{p}\cdot\vec{q} -t_3 q_{0}^2\right]} =  \\
 \frac{d}{d(t_1+t_2)}\frac{d}{dt_3} & & \left\{\frac{1}{(2\pi)^d} \left(\frac{\pi}{t_1+t_2}\right)^{d/2} \frac{\sqrt{t_1 + t_2}}{(t_1 + t_2 + t_3)^{1/2}} e^{\left[\frac{t_{1}^2\vec{p}^2}{(t_1+t_2)} - \frac{t_3 t_{1}^2 p_{0}^2}{(t_1+t_2)(t_1+t_2+t_3)}\right]}\right\} 
\end{eqnarray*}
In order to compute the divergent part of this integral we concentrate on the
terms with no powers of $p$ in the numerator from the differentiation since it
is not hard to convince oneself that all terms of higher order in $p$ converge
in $d=3$.  Thus to extract the $\frac{1}{3-d}$ divergence we have 
\begin{eqnarray*}
&p_{0}&\int\frac{d^3q}{(2\pi)^3}\frac{\vec{q}^2q_{0}^2}{\sqrt{\vec{q}^{2}}
(\vec{p}-\vec{q})^{2}(q_{0}^{2}+\mu^2)^2} \sim \\
-&&\frac{p_0}{2\sqrt{\pi}}\frac{\pi^{d/2}}{(2\pi)^d}\int_{0}^{\infty}dt_1 dt_2 dt_3 \frac{t_3}{\sqrt{t_2}}\left[\frac{\frac{1-d}{2}(t_1+t_2)^{-(1-d)/2}}{(t_1+t_2+t_3)^{3/2}} - \frac{3}{2}\frac{(t_1+t_2)^{(1-d)/2}}{(t_1+t_2+t_3)^{5/2}}\right] \\&& exp\left[\frac{t_{1}^2\vec{p}^2}{(t_1+t_2)} - \frac{t_3 t_{1}^2 p_{0}^2}{(t_1+t_2)(t_1+t_2+t_3)} -t_1\vec{p}^2 -\mu^2 t_3\right]
\end{eqnarray*}
Changing variables to
\begin{displaymath}
t_1 = stu \quad\quad t_2 = st(1-u) \quad\quad t_3 = s(1-t)
\end{displaymath}
we have
\begin{eqnarray*}
-\frac{p_0}{2\sqrt{\pi}}\frac{\pi^{d/2}}{(2\pi)^d}\int_0^1du\int_0^1dt\int_0^\infty ds\quad s^{3-1/2} e^{-s\left[t(1-u)u\vec{p}^2 + (1-t)t u^2 p_{0}^2 + \mu^2(1-t)\right]} \times \\
\left[\frac{1-d}{2} s^{-(2+d/2)}\frac{(1-t)t^{-d/2}}{(1-u)^{1/2}} - \frac{3}{2}s^{-(2+d/2)}t^{1-d/2}\frac{(1-t)}{(1-u)^{1/2}}\right]
\end{eqnarray*}
The integral over $s$ yields 
\begin{eqnarray*}
&-&\frac{p_0}{2\sqrt{\pi}}\frac{\pi^{d/2}}{(2\pi)^d}\Gamma\left(\frac{3-d}{2}\right)\int_0^1du\int_0^1dt \left[\frac{1-d}{2}(1-t)t^{-d/2}(1-u)^{-1/2} - \right.\\
&-& \left.\frac{3}{2}t^{1-d/2}(1-t)(1-u)^{-1/2}\right] \left\{t(1-u)u\vec{p}^2 + (1-t)t u^2 p_{0}^2 + \mu^2(1-t)\right\}^{(d-3)/2} 
\end{eqnarray*}
Expanding the term $\{\cdots\}^{(d-3)/2}$ for $d \rightarrow 3$ and
integrating over the remaining two parameters we finally obtain
\begin{displaymath}
\int\frac{d^3q}{(2\pi)^3}\frac{p_0\vec{q}^2}{\sqrt{\vec{q}^{2}}
(\vec{p}-\vec{q})^{2}q_{0}^{2}}|_{div} = -\frac{p_0}{2\pi^2}\frac{1}{3-d}
\end{displaymath}
In a similar fashion we arrive at
\begin{displaymath}
\int\frac{d^3q}{(2\pi)^3}\frac{q_0\vec{q}^2}{\sqrt{\vec{q}^{2}}
(\vec{p}-\vec{q})^{2}q_{0}^{2}}|_{div} = \frac{p_0}{\pi^2}\frac{1}{3-d}
\end{displaymath}
and thus finally
\begin{eqnarray}\label{sigma0}
\frac{N}{8}i\Sigma_0|_{div} &=& \gamma_0 \left[\underbrace{\frac{3p_0}{2\pi^2}\frac{1}{3-d}}_{\mathbb{A}} - \underbrace{\frac{p_0}{6\pi^2}\frac{1}{3-d}}_{\mathbb{B}} \right] \nonumber \\
 &=& \frac{4}{3\pi^2}\gamma^0 p_0 \frac{1}{3-d}
\end{eqnarray}
The spatial components of the self energy can be evaluated similarly to give
\begin{equation}\label{sigmax}
\frac{N}{8}i\Sigma_{\mathbf{x}} = \frac{2}{3\pi^2}\sum_{i=x,y}\gamma^i p_{i} \frac{1}{3-d}
\end{equation}
where the following results were used
\begin{eqnarray*}
\int\frac{d^3q}{(2\pi)^3} \frac{q_{k}q_{j}}{\sqrt{\vec{q}^{2}}
(\vec{p}-\vec{q})^{2}q_{0}^{2}}|_{div} &=& -\frac{\delta_{j k}}{2\pi^2}\frac{1}{3-d} \\
\int\frac{d^3q}{(2\pi)^3} \frac{q_j \vec{q}^2}{\sqrt{\vec{q}^{2}}
(\vec{p}-\vec{q})^{2}q_{0}^{2}}|_{div} &=& -\frac{p_j}{3\pi^2}\frac{1}{3-d} \\
\int\frac{d^3q}{(2\pi)^3} \frac{\vec{q}^2}{\sqrt{\vec{q}^{2}}
(\vec{p}-\vec{q})^{2}q_{0}^{2}}|_{div} &=& -\frac{1}{2\pi^2}\frac{1}{3-d}
\end{eqnarray*}
with $j,k = 1,2$ the spatial components 

\section{Appendix B: Static boson self energy}\label{appB}

To calculate the static boson self energy we need to evaluate
\begin{equation}\label{boson_sigma1}
-\Sigma_{b}(p_0) = \int\frac{d^3q}{(2\pi)^2}\frac{i\gamma_{0}D_{00}(\vec{q})i\gamma_{0}}{i(p_{0}-q_{0})}
\end{equation}
where $D_{00} = \frac{8}{N}\frac{\mathbf{q}^2}{\vec{q}^3}$ in the Landau
gauge. To combine the denominators we use the following variant of Feynman's
trick
\begin{displaymath}
\frac{1}{a^\alpha b^\beta} = \frac{\Gamma(\alpha + \beta)}{\Gamma(\alpha)\Gamma(\beta)}\int_0^{\infty}\frac{y^{\beta-1}dy}{(a+yb)^{\alpha+\beta}}
\end{displaymath}
which in our case gives
\begin{displaymath}
\frac{1}{|\vec{q}|^3(p_0-q_0)} = \frac{3}{2}\int_{0}^{\infty}\frac{dy}{(\vec{q}^2 + y(p_0-q_0))^{5/2}}
\end{displaymath}
Combining the denominator $(\vec{q} - y/2\vec{v})^2 - y^2/4 + yp_0$ with
$\vec{v}=(1,0,0)$ and shifting the momentum to $\vec{k} = \vec{q} -
y/2\vec{v}$ we obtain
\begin{displaymath}
i\frac{N}{8}\Sigma_{b}(p_0) = \frac{3}{2}\int_{0}^{\infty} dy \int\frac{d^dk}{(2\pi)^d}\frac{\mathbf{k}^2}{[\vec{k}^2 + \Delta]^{5/2}}
\end{displaymath}
which with 
\begin{eqnarray*}
\int\frac{d^dk}{(2\pi)^d}\frac{k_{\mu}k_{\nu}}{[\vec{k}^2 + \Delta]^{5/2}} &=&\frac{1}{2}\frac{\delta_{\mu\nu}}{(4\pi)^{3/2}}\frac{\Gamma(\frac{3-d}{2})}{\Gamma(\frac{5}{2})} \Delta^{\frac{d-3}{2}} \\
\Delta &=& yp_0 - \frac{y^2}{4}
\end{eqnarray*}

The final parameter integral is calculated using \cite{Grozin}
\begin{displaymath}
\int_0^{\infty} y^\alpha(ay+b)^\beta dy = \frac{b^{\alpha+\beta+1}}{a^{\alpha+1}}\frac{\Gamma(-1-\alpha-\beta)\Gamma(1+\alpha)}{\Gamma(-\beta)} 
\end{displaymath}
and yields
\begin{displaymath}
\int_{0}^{\infty}dy(yp_0-\frac{y^2}{4})^{\frac{d-3}{3}} = \frac{p_0}{\frac{1}{4}} \frac{\Gamma(3-d)}{\Gamma(\frac{3-d}{2})}
\end{displaymath}
which finally gives
\begin{equation}
\Sigma_{b}(p_0) = -i \frac{8}{N}\frac{p_0}{\pi^2}\frac{1}{3-d}
\end{equation}

\section{Appendix C: FT of the hole Green's function}\label{appC}

In this appendix we analyze
\begin{equation}\label{G_start_appendC}
G^{e}({\bf Q}_1+\vec{p},\omega_n) = \int d^2r \int_0^\infty d\tau e^{-i \vec{p}\cdot \vec{r}}e^{-[\om]\tau}\frac{m_b L^{-\alpha}}{2\pi\tau}e^{-\frac{m_b r^2}{2\tau}} \frac{\left[ \tau - i \frac{x+y}{\sqrt{2}v}\right]v}{4\pi(\tau^2v^2 + \vec{r}^2)^{\frac{3-\alpha}{2}}}
\end{equation}
more fully than was done in the main body of this paper. Let us split the
above integral into three parts
\begin{eqnarray*}
\mathbb{I}_{1} &=& \int d^2r \int_0^\infty d\tau e^{-i \vec{p}\cdot \vec{r}}e^{-[\om]\tau}\frac{m_b L^{-\alpha}}{2\pi\tau}e^{-\frac{m_b r^2}{2\tau}} \frac{ \tau v }{4\pi(\tau^2v^2 + \vec{r}^2)^{\frac{3-\alpha}{2}}} \\
\mathbb{I}_{2} &=& \int d^2r \int_0^\infty d\tau e^{-i \vec{p}\cdot \vec{r}}e^{-[\om]\tau}\frac{m_b L^{-\alpha}}{2\pi\tau}e^{-\frac{m_b r^2}{2\tau}} \frac{- i \frac{x}{\sqrt{2}}}{4\pi(\tau^2v^2 + \vec{r}^2)^{\frac{3-\alpha}{2}}} \\
\mathbb{I}_{3} &=& \mathbb{I}_{2} (x\rightarrow y)
\end{eqnarray*}
and first concentrate on $\mathbb{I}_1$ After angular integration we arrive at
\begin{displaymath}
\mathbb{I}_1 = K_1\int_0^\infty d\tau \int_0^{\infty}dr r e^{-[\om]\tau}J_0(pr)\frac{e^{-\frac{m_b r^2}{2\tau}}}{(\tau^2 v^2 + r^2)^{\frac{3-\alpha}{2}}}
\end{displaymath}
with $J_{0}$ the zeroth order Bessel function and $K_1 = \frac{m_b v L^{-\alpha}}{4\pi}$. Next we utilize the series representation of $J_0$ 
\begin{displaymath}
J_{\nu}(x) = \left(\frac{1}{2}x\right)^{\nu} \sum_{k=0}^{\infty} \frac{\left(-\frac{1}{4}x^2\right)^{k}}{k!\Gamma(\nu+k+1)}
\end{displaymath}
which converges absolutely for all $x$. This fact allows us to integrate the series term by term 
\begin{displaymath}
\mathbb{I}_1 = K_1  \sum_{k=0}^{\infty} \frac{\left(-\frac{1}{4}p^2\right)^{k}}{k!\Gamma(k+1)}\int_0^{\infty} d\tau e^{-[\om]\tau} (\tau v)^{\alpha - 3}\int_0^\infty dx x^k e^{-\frac{m_b x}{2\tau}}\left[1 + \frac{x}{\tau^2 v^2}\right]^{\frac{\alpha-3}{2}}
\end{displaymath}
Using GR5 (3.383 (5))
\begin{displaymath}
\int_0^{\infty}e^{-px}x^{q-1}(1+ax)^{-\nu} dx = a^{-q}\Gamma(q) \Psi(q,q+1-\nu;p/a) \quad Re q > 0 \quad Re p > 0 \quad Re a > 0 
\end{displaymath}
where $\Psi(a,b;z)$ denotes a confluent Hypergeometric series. The $\tau$ integration can again be performed with the help of GR5 (7.621 (6))
\begin{eqnarray*}
\int_0^{\infty} t^{b-1}\Psi(a,c;t)e^{-st}dt &=& \frac{\Gamma(b)\Gamma(b-c+1)}{\Gamma(a+b-c+1)} s^{-b}\left. \right._2 F_1(a,b;a+b-c+1;1-s^-1) \\
& &Re b>0 \quad Re c < Re b + 1
\end{eqnarray*}
and the identifications $b = 2k+\alpha \quad a = k+1 \quad c = k + \frac{1+\alpha}{2} \quad s = 2[\om]/(m_b v^2)$ yielding to
\begin{eqnarray}
\mathbb{I}_1 &=& \frac{m_b \Gamma^\alpha}{8\pi}\sum_{k=0}^{\infty} \frac{\left(-\frac{p^2v^2}{4}\right)^k}{k!} [\om]^{-2k-\alpha}\frac{\Gamma(2k+\alpha)\Gamma(k+\frac{1+\alpha}{2})}{\Gamma(2k+\frac{3+\alpha}{2})} \nonumber \\
& &\left. \right._2 F_1\left(k+1,2k+\alpha;2k+\frac{3+\alpha}{2};1-\frac{m_bv^2}{2[\om]}\right) 
\end{eqnarray}
Notice that this result reduces to the one quoted in the main body of the thesis for the case $p = 0$ where only the $k=0$ term contributes in the sum.

Next let's look at $\mathbb{I}_2$ which can be cast into
\begin{displaymath}
\mathbb{I}_2 = \frac{m_b L^{-\alpha}}{4\sqrt{2}\pi}\int_0^{\infty} \frac{d\tau}{\tau} \frac{d}{dp_x}\int_0^{\infty} dx e^{-\frac{m_b x}{2\tau}}(\tau v)^{\alpha-3}[1+\frac{x}{\tau^2 v^2}]^{\frac{\alpha-3}{2}}J_0(p\sqrt{x})
\end{displaymath}
With $\frac{d}{dp_x} J_0(p\sqrt{x}) = - J_1(p\sqrt{x})p_x/p$ the above program
can now be repeated step by step. Thus combining $\mathbb{I}_1,\mathbb{I}_2$
and $\mathbb{I}_3$ we find for the physical hole Green's function
\begin{eqnarray}\label{full_answer}
& &G^{e}({\bf Q}_1+\vec{p},\omega_n) = \\
& & =\frac{m_b \Gamma^\alpha}{8\pi}\sum_{k=0}^{\infty} \frac{\left(-\frac{p^2v^2}{4}\right)^k}{k!} [\om]^{-2k-\alpha}\frac{\Gamma(2k+\alpha)\Gamma(k+\frac{1+\alpha}{2})}{\Gamma(2k+\frac{3+\alpha}{2})} \nonumber \\
& & \quad \quad \quad \left. \right._2 F_1\left(k+1,2k+\alpha;2k+\frac{3+\alpha}{2};1-\frac{m_bv^2}{2[\om]}\right) \nonumber \\
& & -\frac{m_b \Gamma^\alpha}{8\pi}\frac{v(p_x + p_y)}{\sqrt{2}}\sum_{k=0}^{\infty} \frac{\left(-\frac{p^2v^2}{4}\right)^k}{k!} [\om]^{-2k-\alpha-1}\frac{\Gamma(2k+\alpha+1)\Gamma(k+\frac{1+\alpha}{2})}{\Gamma(2k+\frac{5+\alpha}{2})} \nonumber \\
& & \quad \quad  \quad\left. \right._2 F_1\left(k+2,2k+\alpha+1;2k+\frac{5+\alpha}{2};1-\frac{m_bv^2}{2[\om]}\right) \nonumber
\end{eqnarray}

\end{widetext}

\end{document}